\documentclass{aa}
\usepackage{graphicx}
\input{psfig.sty}
\usepackage{layout}
\usepackage{aalongtable}
\usepackage{rotating}
\usepackage{lscape}

\def\ecs{ergs cm$^{-2}$ s$^{-1} \ $}
\def\es{ergs s$^{-1}\ $}

\def\lae{\mathrel{<\kern-1.0em\lower0.9ex\hbox{$\sim$}}}
\def\gae{\mathrel{>\kern-1.0em\lower0.9ex\hbox{$\sim$}}}

\begin{document}


\title{Exploring the X-ray Sky with the
XMM-Newton Bright Serendipitous Survey\thanks{Based on observations obtained with XMM-Newton, an ESA science
mission with instruments and contributions directly funded by ESA Member
States and the USA (NASA). The majority of the new optical
spectroscopy data used here have been obtained using the facilities of the
Italian ``Telescopio Nazionale Galileo" (TNG) and of the European  Southern 
Observatory (ESO).}
}

\author{R. Della Ceca \inst{1},
        T. Maccacaro  \inst{1},
    A. Caccianiga \inst{1},
    P. Severgnini  \inst{1},
    V. Braito     \inst{1},
    X. Barcons    \inst{2},
        F.J. Carrera  \inst{2},
    M.G. Watson   \inst{3},
    J.A. Tedds    \inst{3},
    H. Brunner    \inst{4},
    I. Lehmann    \inst{4},
    M.J. Page     \inst{5},
    G. Lamer      \inst{6} \and
    A. Schwope    \inst{6}
    }

\offprints{R. Della Ceca}

\institute{
INAF-Osservatorio Astronomico di Brera, via Brera 28, I-20121 Milano, Italy.
\email{rdc@brera.mi.astro.it, tommaso@brera.mi.astro.it,
caccia@brera.mi.astro.it, paola@brera.mi.astro.it, braito@brera.mi.astro.it}
\and
Instituto de Fisica de Cantabria (CSIC-UC), Avenida de los Castros,
39005 Santander, Spain. \email{barcons@ifca.unican.es, carreraf@ifca.unican.es}
\and
X-ray Astronomy Group, Department of Physics and Astronomy,
Leicester University, Leicester LE1 7RH, UK. \email{mgw@star.le.ac.uk,
jat@star.le.ac.uk}
\and
Max-Planck-Institut f\"ur Extraterrestrische Physik, Postfach 1312, 85741
Garching, Germany. \email{hbrunnerl@mpe.mpg.de, ile@mpe.mpg.de}
\and
Mullard Space Science Laboratory, University College London, Holmbury St. Mary, Dorking, Surrey
RH5 6NT, UK. \email{mjp@mssl.ucl.ac.uk}
\and
Astrophysikalisches Institut Potsdam (AIP), An der Sternwarte 16,
14482 Potsdam, Germany.\email{glamer@aip.de, aschwope@aip.de}
}

\date{}

\abstract{ We present here ``The XMM-Newton Bright Serendipitous Survey",
composed of two flux-limited samples: the XMM-Newton Bright Source Sample (BSS,
hereafter) and the XMM-Newton ``Hard" Bright Source Sample (HBSS, hereafter)
having a flux limit of $f_x\simeq 7 \times 10^{-14}$ erg cm$^{-2}$ s$^{-1}$ 
in the
0.5-4.5 keV and 4.5-7.5 keV energy band, respectively. After discussing the
main goals of this project and the survey strategy, we present the basic
data on a complete sample of 400 X-ray sources (389 of them belong to
the BSS, 67 to the HBSS with 56 X-ray sources in common) derived from the
analysis of 237 suitable XMM-Newton fields (211 for the HBSS). 
At the flux limit of the survey we cover a survey
area of 28.10 (25.17 for the HBSS) sq. deg.
The extragalactic number-flux
relationships (in the 0.5-4.5 keV and in the 4.5-7.5 keV energy bands) are in
good agreement with previous and new results making us confident about the 
correctness of data selection and analysis. 
Up to now $\sim 71$\% ($\sim 90$\%) of the sources have been spectroscopically
identified making the BSS (HBSS) the sample with the highest number of
identified XMM-Newton sources published so far. 
At the X-ray flux limits of the sources studied here we found
that: a) the optical counterpart in the  majority ($\sim 90$\%) of cases has a
magnitude brighter than the POSS II limit (R $\sim 21^{mag}$); b) the 
majority of the objects identified so far are broad line AGN both in the BSS
and in the HBSS. No obvious trend of the source spectra (as deduced from the
Hardness Ratios analysis) as a function of the count rate is measured and  the
average spectra of the ``extragalactic" population corresponds to a (0.5--4.5
keV)  energy spectral index of $\sim 0.8$ ($\sim 0.64$) for the BSS (HBSS)
sample. Based on the hardness ratios we infer that about 13\% (40\%) of the 
sources in the BSS (HBSS) sample are 
described by an energy spectral index flatter than that of the cosmic X-ray
background. Based on previous X-ray 
spectral results on a small subsample of objects 
we speculate that all these sources  are indeed absorbed AGN with the $N_H$
ranging from a few times 10$^{21}$ up to few times  10$^{23}$ cm$^{-2}$. We do not find
strong evidence that the 4.5-7.5 keV  survey is sampling a completely different
source population if compared with the 0.5-4.5 keV survey; rather we find that,
as expected from the CXB synthesis models, the hard survey is simply picking up
a larger fraction of absorbed AGN. At the flux limit of the HBSS sample we
measure surface densities of optically type 1 and type 2 AGN of $1.63\pm 0.25$ deg$^{-2}$
and $0.83\pm 0.18$ deg$^{-2}$, respectively; optically 
type 2 AGN represent $34\pm 9\%$
of the total AGN population. Finally, we have found a clear separation, in the
hardness ratio diagram and in the (hardness ratio) vs. (X-ray to optical 
flux ratio) diagram, 
between Galactic ``coronal emitting" stars and extragalactic sources. The 
information and ``calibration" reported in this paper will make 
the existing and incoming
XMM-Newton catalogs a unique resource for astrophysical studies.
}

\authorrunning{Della Ceca et al.,}
\titlerunning{The XMM-Newton Bright Serendipitous Survey}
\maketitle

\section{Introduction}

Deep {\it Chandra} and {\it XMM--Newton}  observations (Brandt et al., 2001;
Rosati et al., 2002; Moretti et al., 2003; Hasinger et al., 2001
Alexander et al., 2003) have
recently resolved $> \sim 80$\%  of the 2--10 keV  cosmic
X-ray background (CXB) into
discrete sources down to  $f_{x}\sim 3\times$10$^{-16}$ erg cm$^{-2}$ s$^{-1}$.

The statistical analysis (stacked spectra and hardness ratios)
performed on these faint samples provide information on the
X--ray spectral properties of the sources making up most of the
CXB. The X--ray data are consistent with AGN  being the dominant
contributors of the CXB  (see Brandt et al., 2004 and reference therein)
and, as  inferred by the X--ray colors, a
significant fraction  of these sources have hard, presumably
obscured, X--ray spectra,  in agreement with the predictions of
CXB synthesis models  (see Setti and Woltjer, 1989;
Madau et al., 1994; Comastri et al.,
1995, 2001; Gilli et al. 2001; Ueda et al., 2003).

However the majority of the sources found in these medium to deep
fields  are too faint to provide good X--ray spectral information.
Furthermore, the extremely faint magnitude  of a large number of
their optical counterparts makes the spectroscopic identifications
very difficult, or even impossible, with the present day
 ground--based optical telescopes.

Thus, notwithstanding the remarkable results obtained by reaching very faint
X--ray fluxes, the broad--band physical  properties (e.g. the relationship
between optical absorption and X-ray obscuration and the reason why AGN with
similar X-ray properties have completely different optical  appearance) are not
yet completely understood. A step forward toward the solution of these problems
has been recently obtained by Mainieri et al., 2002; Piconcelli  et al., 2002;
2003; Georgantopoulos et al., 2004;  Caccianiga et al., 2004 and  Perola et
al., 2004 using samples of serendipitous sources for which medium/good quality 
XMM-Newton and optical data are available.

With the aim of complementing the results obtained by medium to deep X-ray
surveys, the XMM-Newton Survey Science Centre \footnote {The XMM-Newton Survey
Science Centre is  an international collaboration involving a consortium of 10
institutions appointed by ESA to help the SOC in developing the software
analysis system, to pipeline process all the XMM-Newton data, and to exploit
the XMM-Newton serendipitous detections, see http://xmmssc-www.star.le.ac.uk}
(SSC) has conceived  the ``XMM-Newton Bright Serendipitous Survey". This survey
comprises two high galactic latitude ($|b| > 20^o$), flux limited samples of
serendipitous XMM-Newton sources: the XMM Bright Source Sample (hereafter BSS)
and the XMM Hard Bright Source Sample (hereafter HBSS) having a flux limit of
$f_x\simeq 7 \times 10^{-14}$ erg cm$^{-2}$ s$^{-1}$ in the 0.5-4.5 keV and
4.5-7.5 keV energy bands, respectively.  In addition to the issues related to
the CXB, where is now largely accepted that X-ray obscured AGNs play a
significant  (and perhaps major) role, the use of the 4.5--7.5 keV energy band
partially reduces the strong bias against absorbed sources which occurs when
selecting at softer energies (or when selecting in the optical domain), and is
therefore fundamental to study the accretion history in the Universe (see e.g.
Fiore et al., 2003). A similar energy selection band (i.e. 5-10 keV) was
pioneered by the {\it Beppo}SAX-HELLAS (Fiore et al., 2001) and the ASCA-SHEEP
(Nandra et al., 2003) surveys.

The well defined criteria (completeness, representativeness,
etc..) of this sample will allow both a detailed study of individual sources
of high interest, and statistical studies of populations.
In particular, the BSS and HBSS samples  will be fundamental  to
complement other medium and deep XMM-Newton and {\it Chandra}
survey programs (having fluxes 10 to 100 times fainter and
covering a smaller area of the sky) and  will provide a larger
baseline for all evolutionary studies. Moreover, the high X--ray
statistics which characterize most of the  sources in the
``XMM-Newton Bright Serendipitous Survey", combined with the
relative  brightness of their optical counterparts, will allow us
to investigate their physical properties in detail. Indeed this
sample is already contributing to the solution of some critical
open (and ``hot") questions like the relationship between optical
absorption and X-ray obscuration (Caccianiga et al., 2004) and the
physical nature of the ``X-ray bright optically normal galaxies"
(Severgnini et al., 2003). Many of these issues are investigated
with difficulty using the fainter X-ray samples because of their
typical poor counts statistics for each source.

The spectroscopic identifications together with the X-ray
(spectral, morphological and variability) parameters will be made
available to the community and can be used to define statistical
identification procedures to select rare and interesting classes
of X-ray  sources, enabling the application of these procedures to
the vast amount of XMM-Newton serendipitous data that will be
accumulated during the lifetime of the  mission \footnote{One of
the responsibilities of the XMM-Newton SSC is the production of
the XMM-Newton Source Catalogue. This catalogue will provide a
rich and unique resource for generating well-defined samples for
specific studies, using the fact that X-ray selection is a highly
efficient way of selecting certain types of objects, like for
instance AGN, clusters of galaxies and active stars. The first
XMM-Newton Serendipitous Source Catalogue (1XMM), released on 2003
April 7, contains source detections drawn from 585 XMM-Newton EPIC
observations and a total of $\sim$ 30000 individual X-ray sources
having a likelihood value above 8 and good quality flag. The
median flux (in the total photon energy band 0.2-12 keV) of the
catalogue sources is $\sim 3\times 10^{-14}$  erg cm$^{-2}$
s$^{-1}$, with $\sim 12\%$ of them having fluxes below $\sim
1\times 10^{-14}$  erg cm$^{-2}$ s$^{-1}$, see 
http://xmmssc-www.star.le.ac.uk/}.

In this paper we discuss the BSS and the HBSS survey strategy,
we present a complete sample of 400 sources
extracted from the analysis of 237 XMM-Newton fields 
and we discuss some
preliminary statistical results based on the spectroscopic
identification done so far.

This paper is organized as follows. In section 2 we discuss the
survey strategy (e.g. energy selection bands, primary selection
camera and criteria for field and source selection), we
present basic information on the XMM-Newton fields used and on the
sources belonging to the BSS and HBSS samples and we discuss
the completeness of the ``XMM-Newton Bright Serendipitous Survey".
In section 3 we discuss the number-flux relationship, the
identification work done so far, the broad-band X-ray spectral
properties of the sample, the position of the sources in the
diagram obtained using the X-ray spectral information (provided by
the hardness ratio) and the X-ray to optical flux ratio as well as
the surface densities (Log(N$>$S)-LogS) of type 1 and type 2 AGN
in the HBSS sample. Finally, the summary and the conclusions are
reported in section 4. In the appendices we  discuss our approach
to evaluate the background quality of the data used and to deal
with the X-ray sources falling close to the gaps between the CCDs
or close to the edge of the CCDs.
Throughout this paper $H_0 = 65$ km s$^{-1}$ Mpc$^{-1}$ and
$\Omega_{\lambda} = 0.7$, $\Omega_{M} = 0.3$ are assumed;
the energy spectral index, $\alpha_E$, quoted in this paper refers 
to a power-law spectral model having $S_E \propto E^{-\alpha_E}$. 

\section {Survey Strategy and Sample(s) Selection}

\subsection{Selection Energy Band(s)}

We have decided to survey the bright X-ray sky in two
complementary energy bands:
the 0.5--4.5 keV and the 4.5--7.5 keV energy bands.

The choice of the 0.5--4.5 keV energy band is mainly motivated  by
the desire to avoid the very soft photons (minimizing
non-uniformities introduced by the different values of  Galactic
absorbing column densities along the line of sight) and by the
need to compromise between a  broad passband (to favor throughput)
and a narrow passband  (to minimize non-uniformities in the
selection function due to  different source spectra). 
Furthermore in the 0.5--4.5 keV band XMM-Newton has
the highest throughput.

The choice of the 4.5--7.5 keV energy band (one of the energy
bands used in the standard pipeline processing system of
the XMM-Newton data) was instead dictated  by the need to study
the composition of the source population (in terms of observed and
intrinsic energy distribution and absorption properties) as a
function of the energy selection band, comparing the sources
selected  in this band with those selected in the softer 0.5--4.5
keV energy range. Moreover this energy band  reduces the strong
bias against absorbed sources which occurs when selecting at
softer energies.

\subsection {Primary Selection Camera}

The source sample has been defined
using the data from the EPIC MOS2 detector only.
The main reasons for this choice are:
\begin{enumerate}

\item unlike the EPIC pn, the EPIC MOS cameras have a 
detector pattern  that simplifies the analysis of the field. For example, in
the case of the EPIC MOS detectors the source target, in the large majority of
the  observations,  is fully contained in the central chip;

\item the PSF in the 2 EPIC MOSs is narrower than in the EPIC pn. In
particular, the EPIC MOS2 has the ``best"  PSF (FWHM$\sim$4.4$^{\prime\prime}$
and HEW$\sim$13.0$^{\prime\prime}$ at 1.5~keV, see Ehle et al., 2003). As a
comparison, the EPIC pn PSF has FWHM$\sim$6.6$^{\prime\prime}$ and
HEW$\sim$15.2$^{\prime\prime}$, while the EPIC MOS1 PSF has
FWHM$\sim$4.3$^{\prime\prime}$ and HEW$\sim$13.8$^{\prime\prime}$.

\item the gaps between the EPIC MOS chips are narrower than the gaps in the
EPIC pn detector, simplifying source detection and analysis and maximizing the
survey area;

\item unlike with the EPIC pn camera, we can still use part of the EPIC MOS2
observations in large- and small-window mode by only excluding the area
occupied by the central chip. Since $\sim$ 25\% of the observations have been
performed in window mode, retaining these observations  will maximize the
searched area, speeding up the creation and definition of the source sample.

\end{enumerate}

The major disadvantage of the EPIC MOS2 camera when compared to the EPIC pn
camera is the reduced sensitivity, because of its smaller effective area.
However, since the BSS and HBSS samples contain relatively bright sources, and
considering the  minimum exposure times used here (see section 2.5) this lower
efficiency does not affect the source selection of the samples presented here.
Obviously, once a source is detected and included in the sample, additional
information using data from the EPIC MOS1 and pn detectors are collected to
increase the statistics for the X-ray spectra, timing and morphology analysis.

\subsection {Source Detection}

Each EPIC MOS2 observation used here (see Table 2) has been processed through
the pipeline processing system used for the production of the XMM-Newton
Serendipitous Source Catalogue, based on tasks from the XMM-Newton Science
Analysis Software. Full details about the processing system, the pipeline
products as well as the source searching procedures, flux measurements, source 
likelihood parameter, corrections for vignetting and PSF, etc.. can be found in
http://xmmssc-www.star.le.ac.uk. We note that the count rate(s) reported in
this paper have been already corrected for vignetting and PSF.

\subsection {Criteria for Source Selection}

Since the BSS and HBSS samples have been designed to contain
relatively bright X-ray sources not all the sources detected in
each individual MOS2 field are adequate to be included in these samples.
We discuss here the criteria for the BSS and HBSS source selection
within each EPIC MOS2 field:

\begin{enumerate}

\item

{\it BSS sample}:  0.5--4.5 keV count-rate $\geq 1\times 10^{-2}$
cts/s. At this count rate limit, and given the considered range of
MOS2  exposure times (see Table 2 and figure 1), all the selected
sources have a likelihood parameter in the 0.5--4.5 keV energy
band greater than $\sim 18$ (corresponding to a probability for a
random Poissonian fluctuation to have caused the observed source
counts of $1.5\times 10^{-8}$). No further constraint is thus
needed to ensure the source reliability.

{\it HBSS sample}:  4.5--7.5 keV count-rate $\geq 2\times 10^{-3}$
cts/s and likelihood parameter in the  4.5--7.5 keV energy band
greater than 12 (corresponding to a probability of $6\times
10^{-6}$ for a spurious detection).

The combination of count-rate limit(s) and likelihood parameter(s)
of the sources in the BSS and HBSS samples is such that none of
them are expected to be spurious.

The count rate to flux conversion factors (CR2F) depend on the
source spectra and in Table 1 we report some CR2F as a function of
the input source spectra for a fixed $N_H$ value of $3\times
10^{20}$ cm$^{-2}$ (corresponding to the median value for the
XMM-Newton fields used here).

\begin{table}
\begin{center}
\caption{MOS2 Count Rate to Flux Conversion Factors}
\begin{tabular}{rrlllllll}
\hline \hline
$\alpha_E$      & HR2       & Flux (0.5-4.5 keV)   & Flux (4.5-7.5 keV) \\
              &           & $10^{-12}$ \ecs      & $10^{-11}$ \ecs    \\
(1)           & (2)       & (3)                  & (4)                \\
\hline
$-$3.0         & 0.77      & 14.3 ; 14.4 ; 15.0   &  4.53; 4.54; 4.59   \\
$-$2.0         & 0.56      & 12.8 ; 12.9 ; 13.6   &  4.24; 4.25; 4.30    \\
$-$1.0         & 0.22      & 10.6 ; 10.8 ; 11.7   &  3.97; 3.97; 4.03    \\
$-$0.5         & 0.01      &  9.47;  9.61; 10.6   &  3.84; 3.84; 3.90    \\
 0.0           & $-$0.22   &  8.35;  8.49;  9.55  &  3.71; 3.72; 3.77     \\
 0.4           & $-$0.38   &  7.56;  7.70;  8.82  &  3.62; 3.62; 3.68     \\
 0.5           & $-$0.42   &  7.38;  7.53;  8.66  &  3.59; 3.60; 3.65     \\
 0.7           & $-$0.49   &  7.06;  7.20;  8.37  &  3.55; 3.55; 3.61     \\
 0.9           & $-$0.57   &  6.78;  6.92;  8.11  &  3.50; 3.51; 3.56     \\
 1.0           & $-$0.60   &  6.65;  6.79;  8.00  &  3.48; 3.49; 3.54     \\
 1.5           & $-$0.74   &  6.16;  6.31;  7.61  &  3.38; 3.38; 3.43     \\
 2.0           & $-$0.83   &  5.89;  6.04;  7.46  &  3.28; 3.28; 3.33     \\
 3.0           & $-$0.94   &  5.75;  5.94;  7.62  &  3.10; 3.10; 3.15     \\
\hline \hline
\end{tabular}
\end{center}

Columns are as follows: (1) Energy spectral index
($S_E \propto E^{-\alpha_E}$); (2) corresponding hardness
ratio HR2 computed as discussed in
section 3.3; (3) Flux (corrected for Galactic absorption) in the
0.5-4.5 keV energy band corresponding to an observed count rate of
1 cts/s in the same energy band. The three numbers refer to thin,
medium and thick filters; (4) Flux (corrected for Galactic
absorption) in the 4.5-7.5 keV energy band corresponding to an
observed count rate of 1 cts/s in the same energy band. The three
numbers refer to thin, medium and thick filters. 
NOTE -- A Galactic absorbing column density of $3\times 10^{20}$ cm$^{-2}$,
the median value for the XMM-Newton fields used here, has been
assumed. Given the range of the Galactic absorbing column density along the
line of sight (from $\sim 5\times 10^{19}$ to $10^{21}$ cm$^{-2}$)
the  CR2F in the 0.5-4.5 keV energy range are accurate to $\pm 18$\%; 
the CR2F in the 4.5-7.5 keV energy range  are independent of
the Galactic $N_H$. 
Please note that in the case of sources having extreme HR2 values 
the observed spectra could be much more complex than a simple power law; 
for these sources the conversion factor reported in the table should 
be considered only
 as indicative and a proper X-ray spectral analysis is needed.

\end{table}

For a source with a power-law spectrum with
energy spectral index $\alpha_E$ between 0.7 and 0.8 the count
rate limit in the two chosen bands corresponds to a flux limit of
$\sim 7\times$10$^{-14}$ erg s$^{-1}$ cm$^{-2}$.

\item sources with a distance from the EPIC MOS2 center between an
inner radius ($R_{in}$) and an outer radius ($R_{out}$).

$R_{in}$ depends  on  the actual size and brightness of the target
and on the window mode.  $R_{in}$ ranges between 0 (e.g. survey
fields with no ``target") and 8 arcmin (e.g. bright/extended X-ray
sources or large- and small-window mode). In this way the area of
the detector ``obscured'' by the presence of the target or not
exposed is excluded from the analysis. To guarantee that all the
sources in the catalogue are truly serendipitous, the size of
$R_{in}$ has been adapted in order to exclude the target and the
sources physically related to the target. $R_{out}$ is, for the
large majority of the fields, equal to  13 arcmin. In the few
overlapping fields  we have excluded  from the analysis the outer
region of one of the overlapping fields in order  to obtain a
mosaic of separate and independent regions on the sky. The values
of $R_{in}$ and $R_{out}$ used for each MOS2 image are listed in
Table 2.

\item we have also excluded the sources too close to the edges of
the  field of view  or to the gaps between the CCDs. These sources
could have either the flux and/or the  source centroid poorly
determined (due to the proximity to the edges and/or the gaps),
and therefore could degrade the quality of the data, and would
require uncertainty corrections thus representing a problem in the
subsequent analysis and interpretation of the data.  In Appendix A
we discuss the procedure used to take into account this problem in
an objective way. Obviously,  the excluded area has been taken
into account in the computation of the sky  coverage.

\end{enumerate}

\subsection {Criteria for Field Selection}

Not all the available EPIC MOS2 pointings are adequate for
producing the BSS and HBSS samples. We have defined a set of
selection criteria to avoid problematic regions of the sky,  to
maximize the availability of ancillary information at other
frequencies (i.e. optical and radio) and to  speed up the optical
identification process.   The majority of the fields selection
criteria are common to the BSS and HBSS; however, as
discussed below, we have been more conservative on the minimum
exposure time and background properties for the fields used to
define the HBSS sample. The criteria adopted for field selection are:

\begin{enumerate}

\item availability to SSC before March 2003 (XMM-Newton fields
that are public or with PI granted permission);

\item  high Galactic latitude ($|b| \geq$20$^{\circ}$) to avoid crowded fields,
to obtain a relatively ``clean" extragalactic sample
and to have magnitude information for the optical counterparts
from the Digital Sky Survey material
(the Automated Plate Machine - APM - catalogue\footnote{http://www.ast.cam.ac.uk/~apmcat/}
is almost
complete for  $|b| \geq$20$^{\circ}$);

\item Galactic absorbing column density along the line of sight
less than 10$^{21}$ cm$^{-2}$, to minimize non-uniformities
introduced by large values of the Galactic $N_H$;

\item exclusion of fields centered on bright and/or extended X-ray or optical
targets and those containing very bright stars in the optical band. In the
first two cases the effective area of sky covered and the actual flux limit are
difficult to estimate correctly, making the derivation of the sky-coverage more
uncertain; in the latter case the search for the optical counterpart of the
X-ray sources could be very difficult or even impossible due to the presence of
the bright star;

\item exclusion of fields south of DEC=$-$80 deg since
it could be very difficult to obtain good quality spectroscopy
given the location of the optical facilities available to us;

\item good-time interval \footnote{The good-time interval is
defined as the on-axis exposure time taken from the exposure map
produced in the XMM pipeline processing system.}
exposure $>\sim$ 5 ks for the BSS and $\geq$ 7 ks for the HBSS.
According to the results presented and discussed below, with these
constraints all the sources in the two samples are detectable
across the whole field of view considered, ensuring a ``flat"
sensitivity and therefore a flat sky coverage at the sampled
fluxes.

\item finally, we have also excluded EPIC MOS2 pointings
suffering from a high background rate (i.e.
accumulated during particle background flares).
The background restriction has been more conservative
for the set of fields that have been used to define the
HBSS sample since the overall background is more critical
given the faintness of the sources in the 4.5--7.5 keV energy
band. We have defined and computed
in an automatic way a Background Estimator Parameter (see Appendix B)
which is roughly proportional to the
``real background" in the MOS2 images used.
The set of fields that have been used to define
the HBSS sample must have the  Background Estimator
Parameter less than 100.

\end{enumerate}

Note that we have also considered the EPIC MOS2 observations in
large- and small-window mode satisfying the criteria discussed
above; in these cases we have excluded from the analysis a circular area
of 8 arcmin radius enclosing  the central chip.
No restrictions on the blocking filter in front of the
MOS2 camera have been applied since, as shown in section
2.6, the filter used does not affect the statistical properties of
the sample
\footnote{
The fraction of MOS2 images with a thin, medium and 
thick filter used here are 
$\sim 48$\%, $\sim 46$\% and $\sim 6$\%, respectively}

The complete BSS sample reported here is based on the analysis of
237  XMM-Newton fields, while the complete HBSS sample is based on
a ``restricted" data set of 211 XMM-Newton pointings.

In Table 2 we report basic information on the XMM-Newton MOS2
fields used for the sample selection; in particular we list
the XMM-Newton observation number, the blocking
filter in front of the
MOS2 instrument, the Right ascension and Declination of the MOS2
image center, the on-axis good-time exposure for the MOS2 detector, the
logarithm of the Galactic Hydrogen column density along the line
of sight (from Dickey and Lockman, 1990),
the inner and outer radius of the part of the MOS2
image used in the survey, and the total number of BSS and HBSS sources
found in the surveyed area of each MOS2 image.
In Table 2 we have also marked the 26 MOS2 fields not used for the production 
of the HBSS sample.

In Figure 1 we show the distribution of the MOS2 on-axis good-time exposure,
the distribution of the Galactic hydrogen column density
along the line of sight and
the distribution of the Background
Estimator Parameter for the XMM-Newton MOS2 data-set used.

\begin{table*}
\begin{center}
\caption{Basic information on the XMM-Newton MOS2 fields used for the 
sample selection. We have reported here just few lines of Table 2 for 
illustration purpose. The complete Table 2 will be published only in the 
electronic version of the Journal.}
\begin{tabular}{lllrlllll}
\hline
\hline
Obs. ID         & Filter  & RA; DEC (J2000)        & Exposure       & Log Nh    & $R_{in}$ & $R_{out}$ & BSS  & HBSS  \\
                &         &                        & s              & cm$^{-2}$ & arcmin   & arcmin    & srcs    & srcs    \\
(1)             & (2)     & (3)                    & (4)            & (5)       & (6)      & (7)       &  (8)     &  (9)      \\
\hline
0125310101 & Medium & 00 00 30.4 $-$25 06 43.4 &    19162.6 &         20.27 &          1 &         13 &  4 &  0 \\
0101040101 & Medium & 00 06 19.7 +20 12 22.8 &    34044.7 &         20.60 &          8 &         13 &  1 &  0 \\
0127110201 & Thin1  & 00 10 31.2 +10 58 40.7 &     7558.0 &         20.76 &          2 &         13 &  2 &  0 \\
...        & ...    & ...                    & ...        & ...           & ...        & ...        & ... & ... \\   
...        & ...    & ...                    & ...        & ...           & ...        & ...        & ... & ... \\   
\hline \hline
\end{tabular}
\end{center}
Columns are as follows: (1) XMM-Newton Observation number; (2)
Filter used; (3) Right ascension and Declination (J2000) of the
MOS2 image center; (4) On-Axis good-time exposure; (5) Logarithm of the
Galactic hydrogen column density along the line of sight from
Dickey and Lockman, 1990; (6)
Inner radius of the part of the MOS2 image used; 
(7) Outer radius of the part of the MOS2 image used;
(8) total number of BSS sources found in the surveyed area of each MOS2 image;
(9) total number of HBSS sources found in the surveyed area of each MOS2 image. 
See section 2.4 for details.
\end{table*}

\subsection {The XMM-Newton BSS and HBSS samples.}

Applying the source selection criteria discussed in section 2.4 to
the MOS2 fields reported in Table 2 we have selected 400
XMM-Newton sources: 389 sources belong to the BSS sample and 67
sources to the HBSS  sample with 56 sources in common. Basic
information on the sources are reported in Table 3 (BSS) and in
Table 4 (HBSS); in particular we report the source name, the
XMM-Newton observation number, Right Ascension and Declination
(J2000) of the X-ray source position, the angular distance (in
arcmin) between the source and the MOS2 image center, the source
count rate in the 0.5--4.5 keV energy band (BSS sample) or in the
4.5--7.5 keV energy band (HBSS sample), the hardness ratios
computed as described in section 3.3, and the optical
spectroscopic classification (see section 3.2 for details). In
Table 4 we have also marked the 11 sources belonging to the HBSS
sample but not to the BSS sample.

In Figure 2  we show  the surface density of the sources belonging
to the BSS and to the HBSS as a function of: the MOS2 on-axis
good-time exposure (panel a); the Background Estimator Parameter
(panel b); the off-axis angle (panel c); and the blocking filter
in front of the MOS2 detector (panel d). The appropriate area
covered in each bin has been considered and errors have been
computed using Poisson statistic. The dashed lines reported in
Figure 2 correspond to the mean surface density obtained
considering the whole sample. As can be seen there is no
significant trend of the source surface density with respect to
the plotted parameters confirming a flat sensitivity across the
field (i.e. flat sky coverage at the sampled fluxes)). The only
point which seems to be a factor $\sim$ 2.5 above the other is the
bin at the highest exposure time in the BSS sample (see panel a).
This excess is due to 5 sources found in the field 0022740101
(centered on the Lockman hole), the only pointing in the bin
considered; however the error bars are large and so the reported
surface density is not significantly different from the mean
value. The absolute source surface density as a function of the
flux (LogN-LogS) is also in very good agreement with previous and
new measurements (see section 3.1) making us confident of the
correctness of the data analysis and source selection.

\begin{table*}
\begin{center}
\caption{Basic information on the XMM-Newton BSS sample.
We have reported here just few lines of Table 3 for 
illustration purpose. The complete Table 3 will be published only in the 
electronic version of the Journal.}
\begin{tabular}{lllrrrrl}
\hline
\hline
Name                 & Obs. ID    & RA; DEC (J2000)      & OffAxis  & Rate                     & HR2           & HR3            &  Class  \\
XBS...               &            &                      & arcmin     & $\times 10^{-2}$ cts/s    &               &              &      \\
(1)                  & (2)        & (3)                  & (4)       & (5)                       & (6)           & (7)            & (8)  \\
\hline
J000027.7$-$250442 & 0125310101 & 00 00 27.8 $-$25 04 42.6 &  2.10 &$  1.13 \pm   0.08 $&$ -0.48 ^{+  0.06 }_{-  0.06 }$ &$ -0.37 ^{+  0.11 }_{-  0.11 }$& AGN1     \\
J000031.7$-$245502 & 0125310101 & 00 00 31.8 $-$24 55 02.9 & 11.68 &$  1.41 \pm   0.13 $&$ -0.67 ^{+  0.07 }_{-  0.06 }$ &$ -0.37 ^{+  0.17 }_{-  0.17 }$& AGN1     \\
J000100.2$-$250501 & 0125310101 & 00 01 00.2 $-$25 05 01.2 &  6.97 &$  1.38 \pm   0.10 $&$ -0.37 ^{+  0.07 }_{-  0.07 }$ &$ -0.47 ^{+  0.10 }_{-  0.10 }$& AGN1     \\
...                & ...        & ...                      & ...   & ...                & ...                            & ...                           & ...      \\   
...                & ...        & ...                      & ...   & ...                & ...                            & ...                           & ...      \\   
\hline \hline
\end{tabular}
\end{center}
Columns are as follows: (1)
Source name; (2) XMM-Newton Observation number; (3) Right Ascension and
Declination (J2000) of the source (X-ray position); (4) Angular
distance (in arcmin) between the source and the MOS2 image center;
(5) Source count rate, and $1\sigma$ error, in the 0.5-4.5 keV
energy band (units of $10^{-2}$ cts/s). In Table 1
we have reported the MOS2 conversion factors between the 0.5-4.5
keV count rate and the flux as a function of the energy spectral
index, the hardness ratio HR2 and the blocking filter; (6) and (7)
Hardness ratios computed as described in section 3.3. The errors
on the hardness ratios have been evaluated using simulations and
correspond to $1\sigma$; (8) Optical spectroscopic classification
(AGN1: broad line AGN; AGN2: narrow line AGN; GAL: Optically Normal
Galaxy; CL: Cluster of Galaxies; BL: BL Lac Object; star: star;
?: Tentative classification; see section 3.2 for details).
\end{table*}

\begin{table*}
\begin{center}
\caption{Basic information on the XMM-Newton HBSS sample.
We have reported here just few lines of Table 4 for 
illustration purpose. The complete Table 4 will be published only in the 
electronic version of the Journal.}
\begin{tabular}{lllrrrrl}
\hline
\hline
Name                 & Obs. ID    & RA; DEC (J2000)      & OffAxis  & Rate                     & HR2           & HR3            &  Class  \\
XBS...               &            &                      & arcmin     & $\times 10^{-3}$ cts/s    &               &              &      \\
(1)                  & (2)        & (3)                  & (4)       & (5)                       & (6)           & (7)            & (8)  \\
\hline
J002618.5+105019     & 0001930101 & 00 26 18.5 +10 50 19.3 &           9.56 &$  2.35 \pm        0.55 $& -0.53 $^{+  0.04 }_{-  0.04 }$ & -0.67 $^{+  0.06 }_{-  0.06 }$& AGN1     \\
J013240.1$-$133307   & 0084230301 & 01 32 40.1 $-$13 33 07.8 &      11.90 &$  3.23 \pm    0.71 $& -0.02 $^{+  0.10 }_{-  0.10 }$ & -0.37 $^{+  0.11 }_{-  0.11 }$& AGN2     \\
J013944.0$-$674909   & 0032140401 & 01 39 44.0 $-$67 49 09.4 &       2.69 &$  2.05 \pm    0.46 $& -0.50 $^{+  0.07 }_{-  0.07 }$ & -0.43 $^{+  0.13 }_{-  0.13 }$& AGN1?    \\
...                & ...        & ...                      & ...   & ...                & ...                            & ...                           & ...      \\   
...                & ...        & ...                      & ...   & ...                & ...                            & ...                           & ...      \\   
\hline \hline
\end{tabular}
\end{center}
Columns are as follows: (1) Source name;
(2) XMM-Newton Observation number; (3) Right Ascension and Declination
(J2000) of the source (X-ray position); (4) Angular distance (in
arcmin) between the source and the MOS2 image center; (5) Source
count rate, and $1\sigma$ error, in the 4.5-7.5 keV energy band
(units of $10^{-3}$ cts/s). In Table 1 we have
reported the MOS2 conversion factors between the 4.5-7.5 keV count
rate and the flux as a function of the energy spectral index, the
hardness ratio HR2 and the blocking filter; (6) and (7) Hardness
ratios computed as described in section 3.3. The errors on the
hardness ratios have been evaluated using simulations and
correspond to $1\sigma$; (8) Optical spectroscopic classification
(AGN1: broad line AGN; AGN2: narrow line AGN; GAL: Optically Normal
Galaxy; CL: Cluster of Galaxies; BL: BL Lac Object; star: star;
?: Tentative classification; see section 3.2 for details).
\end{table*}

\begin{figure*}
\begin{center}
\begin{tabular}{lll}
\includegraphics[width=0.33\textwidth]{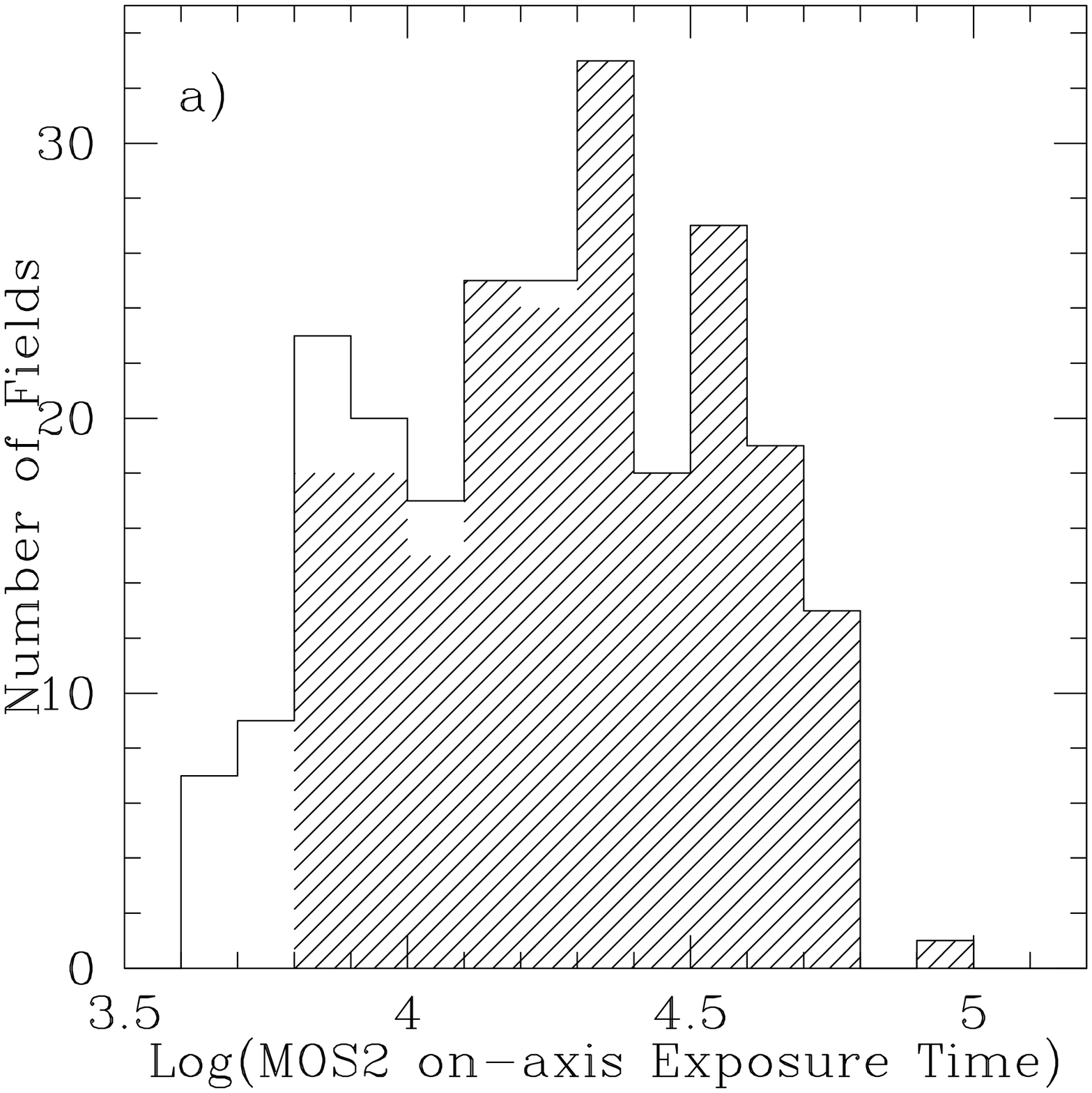}
&\includegraphics[width=0.33\textwidth]{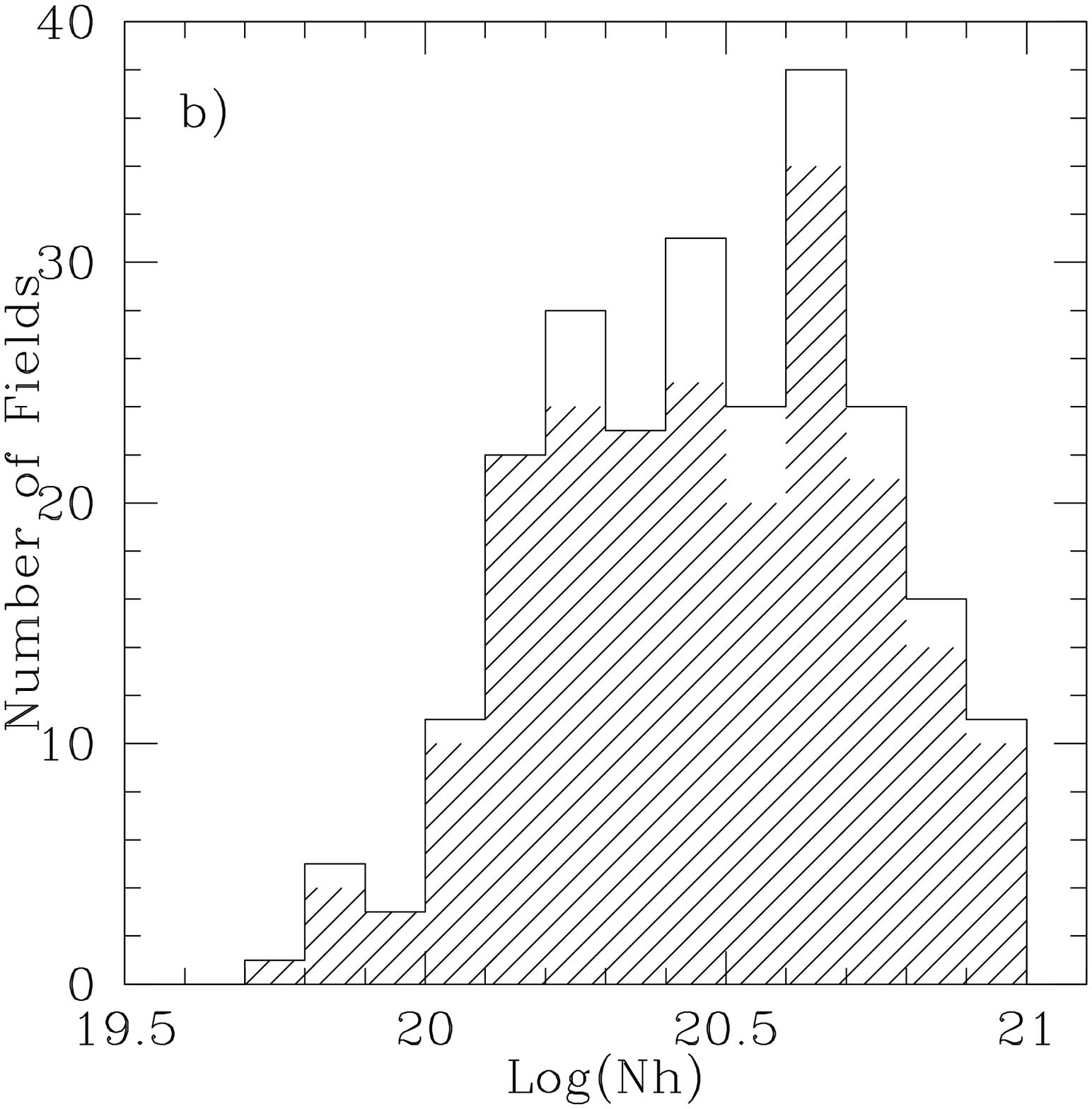}
&\includegraphics[width=0.33\textwidth]{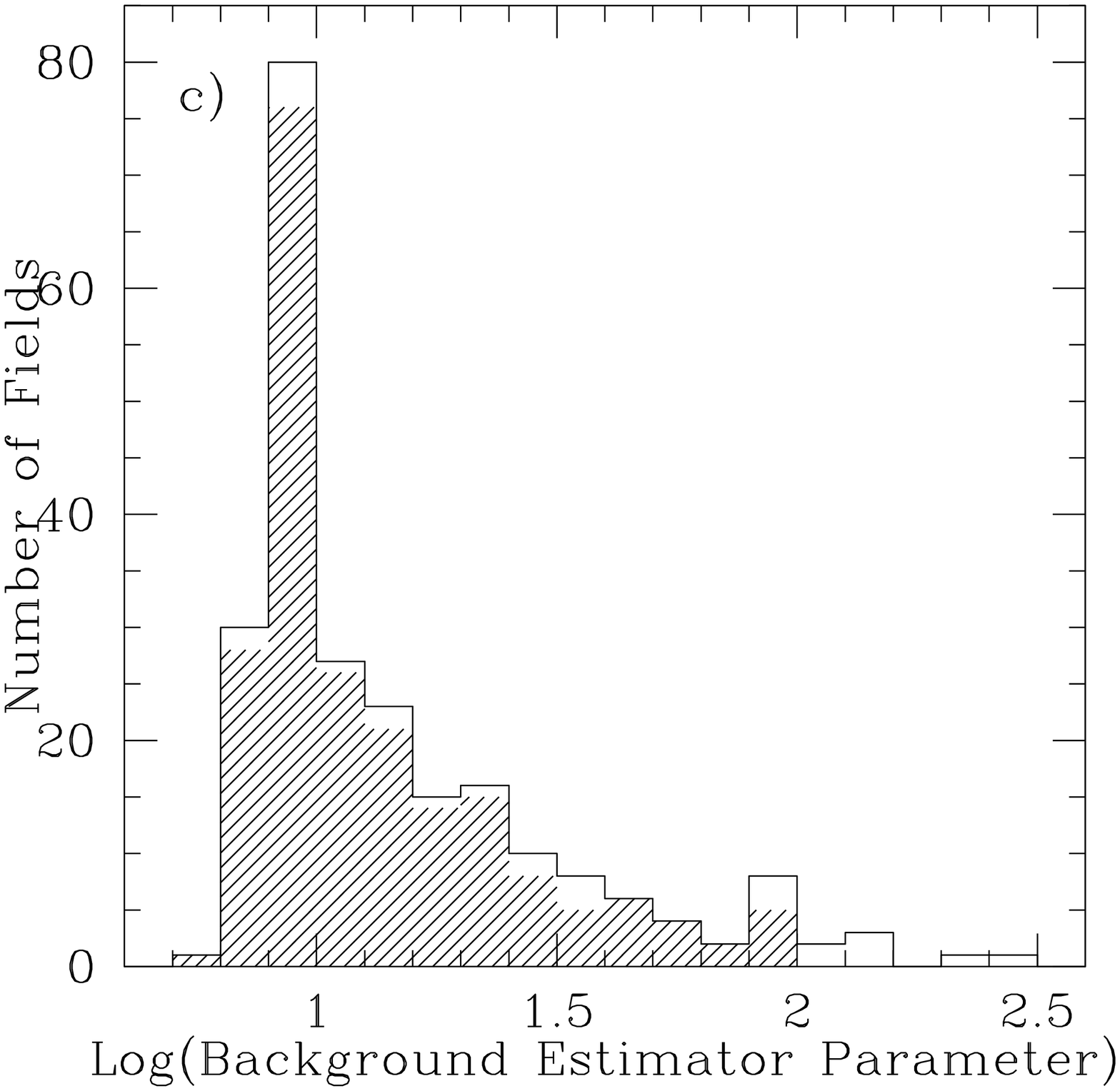}\\
\end{tabular}
\caption{Histograms of some basic properties of the XMM-Newton
MOS2 fields used for the sample selection. Normal histograms refer
to the XMM-Newton fields used to define the BSS sample, while shaded
histograms refer to the subset of 
XMM-Newton fields used to define the HBSS
sample. Panel a: histogram of the MOS2 on-axis good-time exposure;
Panel b: histogram of the Galactic hydrogen column density along
the line of sight; Panel c: histogram of the Background Estimator
Parameter (see Appendix B for details). }
\end{center}
\end{figure*}

\begin{figure*}[htb]
\parbox{10cm}{
\psfig{file=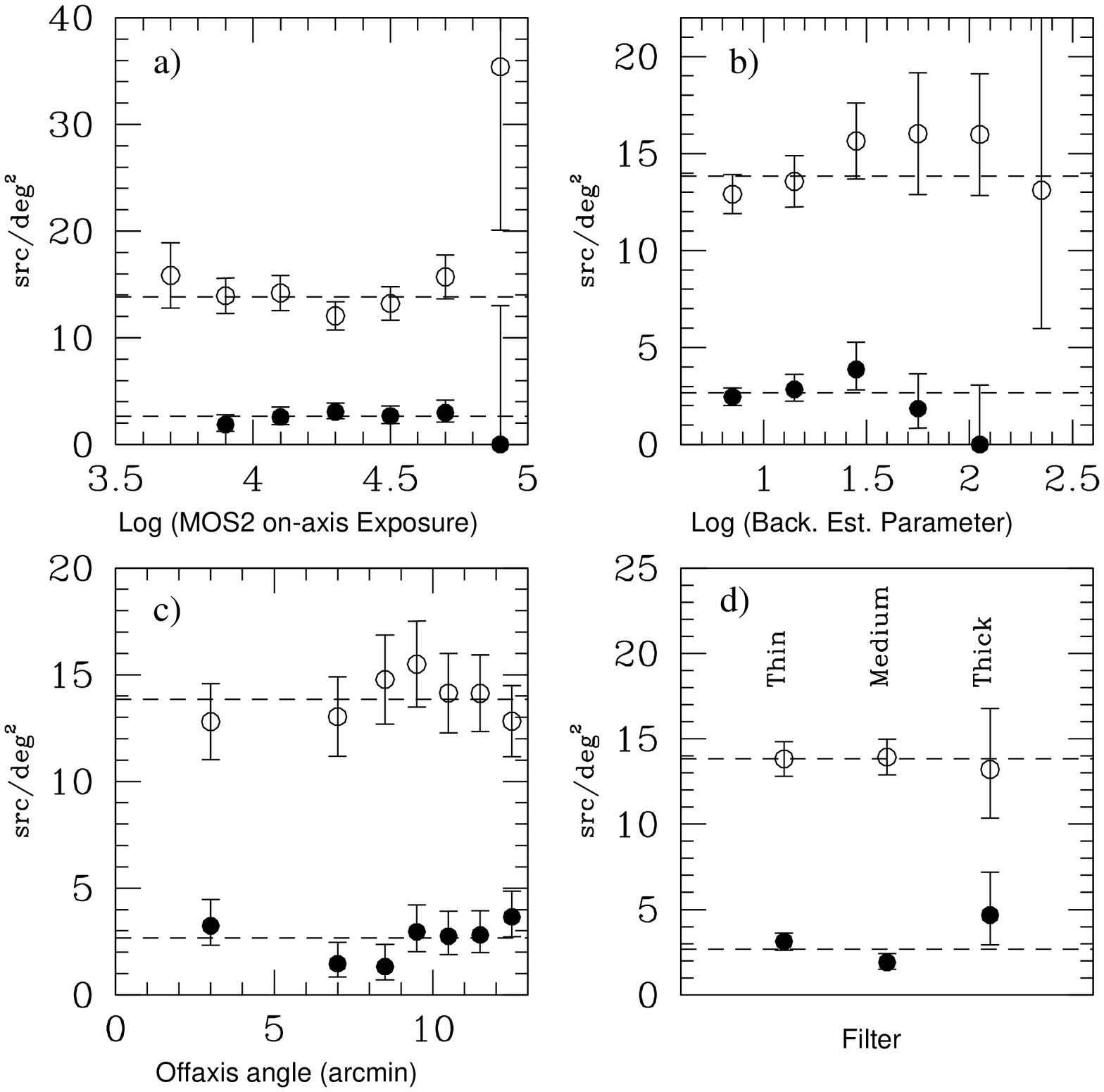,height=18.0cm,width=18.0cm,angle=0} }
\caption{ Panel a: Source surface density as a function of the
MOS2 on-axis good-time exposure for the sources belonging to the BSS
(open circles) and to the  HBSS (filled circles) sample. Panel b:
Source surface density as a function of the Background Estimator
Parameter; symbols as in panel a. Panel c: Source surface density
as a function of the offaxis angle; the bin size has been adapted in order
to have similar areas in each bin; symbols as in panel a. Panel
d: Source surface density as a function of the blocking filter in
front of the MOS2 detector; symbols as in panel a. For all the
panels the dashed lines correspond to the mean surface density
considering the whole BSS or HBSS sample. Errors have been
computed using Poisson statistic.
}
\end{figure*}

\section {First Results}

\subsection {The Number-Counts Relationship(s)}

In Figure 3 we show (filled circles) a binned   representation of
the extragalactic\footnote{Since we are primarily interested in
the extragalactic number-flux relationship  we have excluded from
the computation the sources classified as stars (see section 3.2).
Based on the results presented in section 3.5 we are confident that the large 
majority of the unidentified sources are associated to extragalactic objects.}
number-flux relationships in the 0.5-4.5 keV energy band (panel a) and in the
4.5-7.5 keV energy band (panel b). 
As already shown in section 2.6, the sky coverage of
this survey  at the flux limit used to define the BSS and the HBSS
samples is flat and equal to $\sim 28.10$ sq.deg and $\sim 25.17$ sq.deg, 
respectively; given the flat sky coverage the
errors in the binned representation are Poissonian errors
on the total number of sources having a flux greater than
any fixed flux. A conversion factor appropriate for a  power-law
spectral model with energy index equal to 0.8 (0.7) in the 0.5-4.5
keV (4.5-7.5 keV) energy band, filtered by an $N_{H_{Gal}} \sim 3\times
10^{20}$ cm$^{-2}$ (the median value of the $N_{H_{Gal}}$ of
the survey), has been used in the conversion between the count
rate and the flux. The energy spectral index used in the 0.5-4.5
keV energy band corresponds to the ``average" one of the
extragalactic BSS population in the same energy
selection band (see section 3.3).
In the 4.5-7.5 keV energy band we have used $\alpha_E = 0.7$,  the same energy
spectral index assumed from other recent surveys in the 5-10 keV  band 
(e.g. the HELLAS2XMM survey, Baldi et al., 2002) and very
close  to the median energy spectral index in the 0.5-4.5 keV  band of the
extragalactic HBSS population (see below for details).
We recall
that, given the median $N_{H_{Gal}}$ of the survey, the count rate
to flux conversion factor in the 0.5-4.5 keV (4.5-7.5 keV) energy
band are accurate $\sim \pm 20$ ($\sim \pm 8$)\% for an energy
spectral index in the range between 0 to 2.

\begin{figure*}
\begin{center}
\begin{tabular}{cc}
\includegraphics[width=0.50\textwidth]{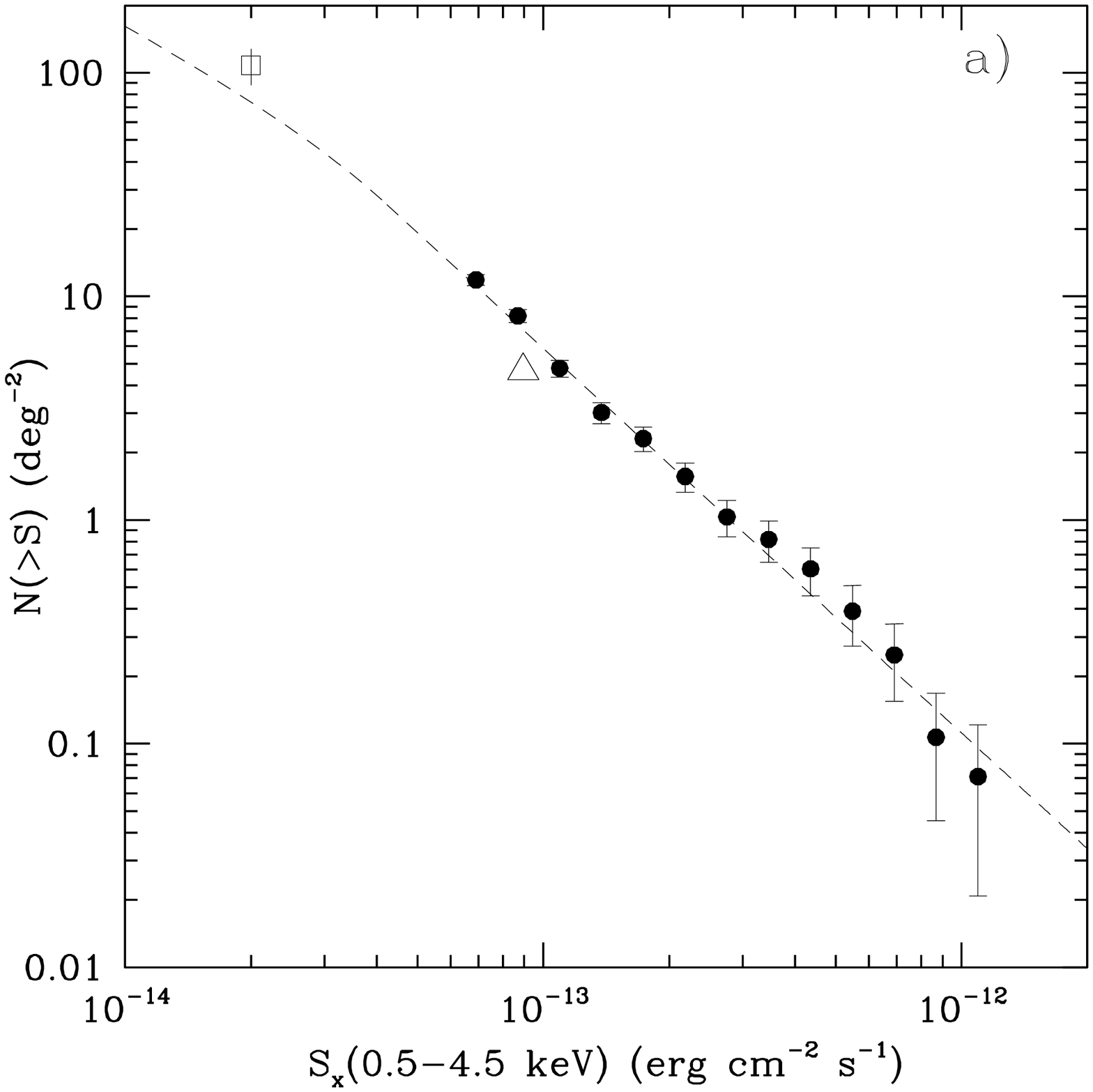}&\includegraphics[width=0.50\textwidth]{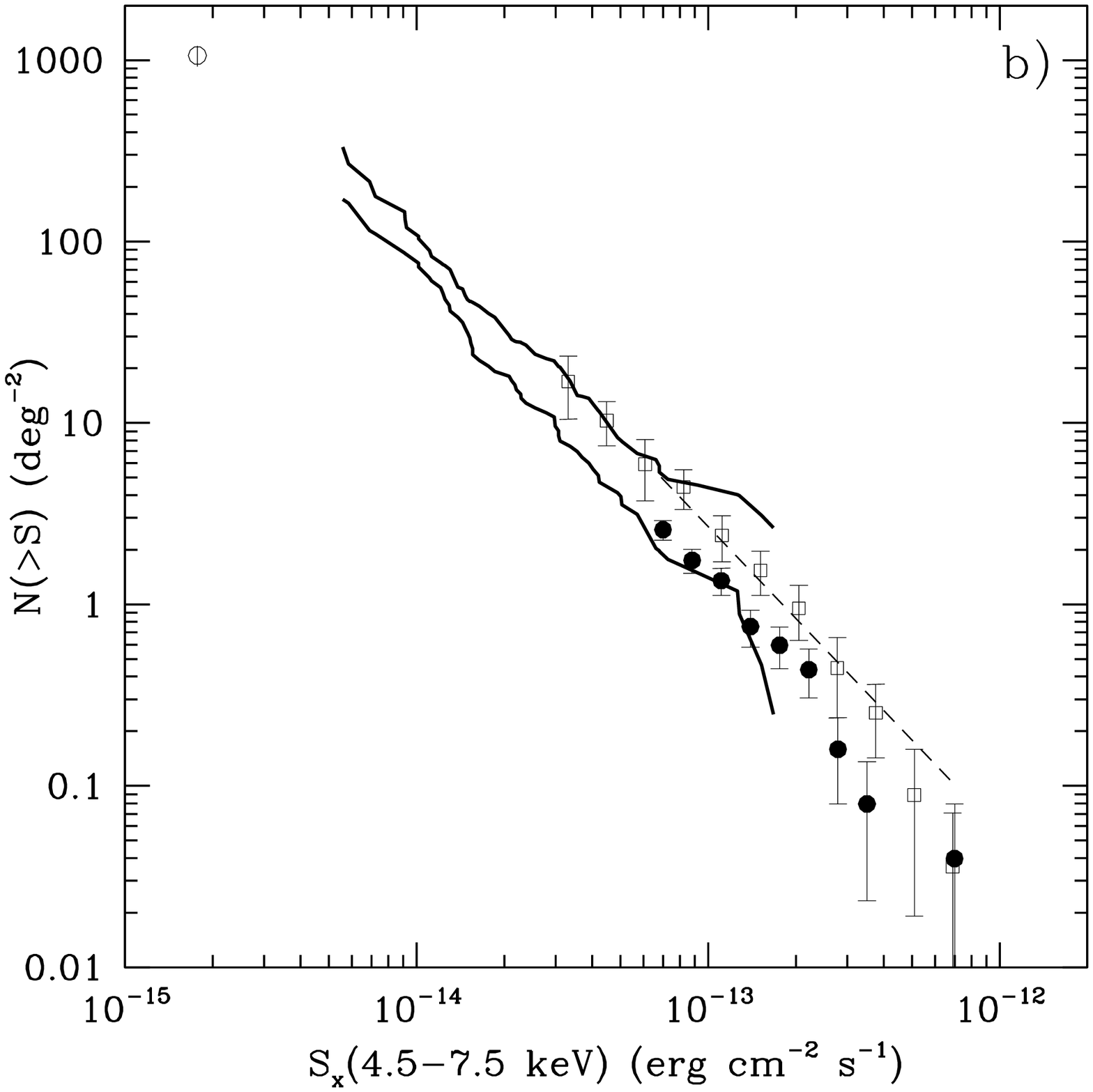}\\
\end{tabular}
\caption{The extragalactic number-flux relationship in the 0.5-4.5
keV energy band (panel a) and in the 4.5-7.5 keV energy band
(panel b) obtained using the BSS and HBSS samples (binned
representation: black filled circles). In the 0.5-4.5 keV
Log(N$>$S)-LogS (panel a) we have also reported the ROSAT (0.5 --
2.0 keV) Log(N$>$S)-LogS (dashed line) and the EMSS (0.3--3.5 keV)
extragalactic number density at $\sim 10^{-13}$ \ecs
(open triangle) both converted to the
0.5--4.5 keV energy band. The open square at $S \simeq 2\times
10^{-14}$ \ecs represents the extragalactic surface density
(0.5-4.5 keV) obtained by the XMM-Newton AXIS Medium Survey
project. In panel b (4.5--7.5 keV Log(N$>$S)-LogS) we have also
reported the HELLAS2XMM Log(N$>$S)-LogS interval (area inside the
thick solid lines), the HELLAS Log(N$>$S)-LogS (open squares) and
the SHEEP Log(N$>$S)-LogS (dashed line). The open circle at $S
\simeq 2\times 10^{-15}$ \ecs represents the extragalactic surface
density (5-10 keV band) in the Lockman hole field. Both these latter number
densities have been converted from their original (5-10 keV) band
to the 4.5--7.5 keV band.}
\end{center}
\end{figure*}

\begin{table}
\begin{center}
\caption{The Extragalactic Log(N$>S$)-LogS maximum likelihood best
fit parameters: 
N($>S$) = K $\times$ (S/10$^{-13}$) $^{-\alpha}$, 
where N($>S$)
is the surface densities of sources having a flux greater than
S in deg$^{-2}$}
\begin{tabular}{lrlllllll}
\hline
\hline
Sample               & Objects        & $\alpha$               & K                \\
                     &                &                        &                   \\
(1)                  & (2)            & (3)                    & (4)                \\
\hline
BSS$^{a}$            & 330            & 1.80$^{1.91}_{1.69}$   & 6.1$^{5.8}_{6.4}$ \\
HBSS                 &  65            & 1.64$^{1.89}_{1.41}$   & 1.5$^{1.4}_{1.6}$ \\
HBSS AGN1            &  41            & 1.72$^{2.03}_{1.42}$   & 0.92$^{0.82}_{1.0}$ \\
HBSS AGN2            &  21            & 1.57$^{1.99}_{1.18}$   & 0.50$^{0.42}_{0.58}$ \\
\hline
\hline
\end{tabular}
\end{center}
Columns are as follows: (1) Sample; (2) Number of
sources used in the fit; (3) Best fit power-law
slope and 68\% confidence intervals; (4) Log(N$>S$)-LogS
Normalization and 68\% confidence intervals.
Note that the normalization K is not a parameter of the
fit but is determined by re-scaling the model to the 
number of objects in the sample.
The normalization K reported here corresponds to the
surface densities at  $1\times 10^{-13}$ \ecs.
NOTE -- 
$^{a}$ As explained  in section 3.1 we have excluded from the fit the 3 
BSS extragalactic sources having a flux greater than  
$1\times 10^{-12}$ \ecs.

\end{table}

Both LogN($>$S)--LogS distributions can be well described by a power-law
model N($>$S) $\propto  S^{-\alpha}$; their best fit spectral
parameters, obtained applying the maximum likelihood method to
the unbinned data (see Maccacaro et al., 1982 for details), are
reported in Table 5.  

The fits have been performed from a flux of $\simeq 7\times
10^{-14}$ \ecs (the faintest flux) to a flux of $\simeq 1\times
10^{-12}$ \ecs (we have excluded from the fit 3 extragalactic 
BSS sources brighter than this flux limit).
For fluxes brighter than this limit we may not be complete since
``bright" X-ray sources were chosen as targets of
observations and then excluded, by definition, from the survey.
However we note that the surface density of sources with flux
greater than $\sim 1\times 10^{-12}$ \ecs 
is such that about 2.7 (0.9)  sources in the BSS (HBSS)
sample are expected given the covered sky area; 
these numbers are fully consistent with what is observed.

Both Log(N$>$S)-LogS derived here have been compared with a number of
representative Log(N$>$S)-LogS reported in the literature. In
particular, in Figure 3 panel a, we have reported: a) the extragalactic
ROSAT (0.5 -- 2.0 keV) Log(N$>$S)-LogS from Hasinger et al., 1998
(dashed line); b) the EMSS (0.3--3.5 keV) extragalactic number
density at $\sim 10^{-13}$ \ecs (open triangle; Gioia et al., 1990)
and c) the extragalactic surface density
obtained from the XMM-Newton AXIS Medium Survey team in the
0.5-4.5 keV energy range (Barcons et al., 2002; open square at $S
\simeq 2\times 10^{-14}$ \ecs). To convert the ROSAT 0.5--2.0
keV band fluxes and the EMSS 0.3-3.5 keV band fluxes
into 0.5--4.5 keV fluxes we have used a power-law
spectral model having $\alpha_E \simeq 1.0$, corresponding to the
mean spectral index of the ROSAT and the EMSS sources
(see Hasinger et al., 1993 and Maccacaro et al., 1988, respectively).

In panel b) we have compared our result with: a) the extragalactic
XMM-Newton (5 -- 10 keV) Log(N$>$S)-LogS from the HELLAS2XMM survey (area
inside the thick solid lines; Baldi et al., 2002); b) the {\it Beppo}SAX-HELLAS (5
-- 10 keV) Log(N$>$S)-LogS (open squares; Fiore et al., 2001); c)
the ASCA-SHEEP (5 -- 10 keV) Log(N$>$S)-LogS (dashed line;  Nandra et al.,
2003). Finally, the open circle at $S \simeq 2\times 10^{-15}$
\ecs represents the extragalactic surface density in the (5 -- 10
keV) energy band from the XMM-Newton observation of the Lockman
hole field (Hasinger et al., 2001). For consistency with previous
hard survey both these latter number densities have been converted
from their original (5-10 keV) band to the 4.5--7.5 keV band using
a power-law spectral model having $\alpha_E = 0.7$.
We found that our results are fully consistent with those obtained
from the other XMM-Newton related survey (e.g. the HELLAS2XMM 5--10 
keV survey) and,
moreover, our better statistics above $7 \times 10^{-14}$
\ecs allow us to significantly constrain the 4.5--7.5 keV extragalactic number
densities above this flux.

On the other hand the extragalactic HBSS Log(N$>$S)-LogS falls below both the
{\it Beppo}SAX-HELLAS and the ASCA-SHEEP determinations. Given the results discussed in
section 2.6 and the very similar slope between our Log(N$>$S)-LogS 
and the {\it Beppo}SAX/ASCA Log(N$>$S)-LogS 
we  have checked if this problem could be
related to an offset of the absolute flux scale in the 4.5--7.5 keV energy
range between XMM-Newton and {\it Beppo}SAX/ASCA.
To this purpose we have cross-correlated the HELLAS and the SHEEP sources with
the total catalogue of XMM-Newton sources obtained from the analysis of the 237
XMM-Newton
fields reported in Table 2.  Using a search radius of 90$^{\prime\prime}$ and
considering the point-like  XMM-Newton sources with a 4.5--7.5 keV likelihood
parameter greater than 12  and with an ``Illumination Factor" (see Appendix A for
details) greater than 0.8,
we have found  6 ``bona fide" HELLAS-XMM coincidences and  2 ``bona fide"
SHEEP-XMM coincidences.  The ratio between the 4.5--7.5 keV XMM-Newton fluxes and the
5-10 keV {\it Beppo}SAX  fluxes in the case of the HELLAS sources ranges
between 0.09 and 0.98 with a  mean value of 0.47, while in the case of the two
SHEEP sources   the ratio between the 4.5--7.5 keV XMM-Newton fluxes and the 5-10 keV
ASCA   fluxes is equal to $\simeq 0.63$ for both objects.    Although 
these small numbers do not allow us to draw firm conclusions we note that using
a conversion factor between the  fluxes in the 5--10 keV energy range and the
fluxes in the  4.5--7.5 keV energy range equal to 0.47 (instead of 0.69 as
expected for $\alpha_E = 0.7$ and as assumed in figure 3) the HBSS
Log(N$>$S)-LogS and the {\it Beppo}SAX-HELLAS Log(N$>$S)-LogS turn  out to be
in perfect agreement. This suggests that an  offset in the  absolute flux scale
could easily explain the disagreement in the number densities discussed above;
this possible discrepancy in the flux scale has to be further
investigated.

\subsection {Optical identification and classification}

One of the main characteristics of the X-ray sources presented
here is that the majority ($\sim$ 90\%) of them have an optical
counterpart above the POSS II limit (R $\sim 21^{mag}$), thus
allowing spectroscopic identification even on 2-4 meter class
telescopes. 
Furthermore, given the good accuracy of the X-ray
positions 
\footnote{Using the optical position of the sources 
classified as Type 1 AGN we have evaluated that the 
90\% confidence level error circle 
has a radius equal to $\sim 4^{\prime\prime}$; 
about 99\% of the Type 1 AGN are within $6^{\prime\prime}$ 
from the X-ray position. This is consistent with what found 
in other XMM-Newton surveys (Barcons et al. 2002; Fiore et al. 
2003)}
and the magnitude of the optical counterparts
there is no ambiguity in the optical identification for the large
majority of cases.

Up to now 285 X-ray sources have been spectroscopically identified
(either from the literature or from our own observations mainly at the
Italian ``Telescopio Nazionale Galileo" -TNG, at the ESO 3.6m, 
at the Calar Alto 2.2 m or at the NOT 2.6m
\footnote{As part of the AXIS (An XMM-Newton
International Programme) project, see {\tt http://www.ifca.unican.es/\~{}xray/AXIS}}
telescopes) leading to a 71\%
and 90\% identification rate for the BSS and HBSS samples
respectively.

The optical breakdown of the sources identified so far is reported
in Table 6. We stress that the source detection algorithm is
optimized for point-like sources, so the sample of clusters of
galaxies is not statistically complete nor representative of the cluster
population.

To our present knowledge all but one (XBS J014100.6$-$675328
\footnote{XBS J014100.6$-$675328 (BL Hyi), also belonging to
the HBSS sample, is a well known AM Herculis object (a polar)
i.e. a binary system composed of a magnetic white dwarf and a low-mass star 
(see Caccianiga et al., 2004 and references therein).}
) of the sources classified as stars are coronal emitters.
\begin{table}
\begin{center}
\caption{The current optical breakdown of the BSS and HBSS
Samples}
\begin{tabular}{lrr}
\hline
\hline
                     &    BSS$^{1}$          & HBSS         \\
                     &                       &              \\
                     &                       &              \\
\hline
Objects$^{2}$        &           389  (166)  &          67  \\
\\
Identified:          &           278  (146)  &          60  \\
Identification rate  &           71\% (88\%) &          90\% \\
                     &                       &              \\
AGN-1                &           180  (100)  &          39  \\
AGN-2                &            26  (15)   &          16  \\
Galaxies$^{3}$       &             7  (3)    &           1  \\
Clusters of Galaxies &             4  (1)    &           1  \\
BL Lacs              &             5  (3)    &           1   \\
Stars$^{4}$          &            56  (24)   &           2   \\
\hline
\hline
\end{tabular}
\end{center}
$^{1}$ In brackets we have reported the optical breakdown for the BSS sources with 
Right Ascension below 5$^h$ or above 17$^h$;
$^{2}$ Note that 56 sources are in common between the BSS and HBSS samples;
$^{3}$ We stress that some of the sources classified as ``Optical Normal
Galaxy" could indeed host an optically elusive AGN (see e.g. Severgnini et al.,
2003);
$^{4}$ To our knowledge all but one (XBS J014100.6$-$675328)
of the sources classified as stars are coronal emitters.
\end{table}
If we consider the BSS sources with Right Ascension below $5^h$ or above $17^h$
(spectroscopic identification rate of $\sim 88$\%, see
Table 6), the X-ray coronal emitting stars represent $\sim 14$\% of the $|b| > 20^o$
(0.5--4.5 keV) population at the sampled fluxes. This fraction must be
compared with $\sim 1.5$\% of coronal emitters in the HBSS  sample
\footnote{On the basis of the results presented in section 3.5 we are
confident that the bulk of unidentified sources both in the  BSS and HBSS
samples are associated with extragalactic objects and we know that XBS
J014100.6$-$675328 is an accreting binary system 
(see Caccianiga et al., 2004).};
this smaller fraction of stars in the HBSS sample, compared with that
in the BSS sample, is  entirely consistent with their low temperature
coronal emission.
Note that in the softer (0.5-2.0 keV) ROSAT Bright Survey Catalog
(RBS, Schwope et al., 2000)
the fraction of coronal emitting stars is around 37\%.

The large majority ($\sim 90$ \%)
of the extragalactic X-ray sources are emission
line objects, i.e.  sources for which at least one strong (EW $>>$
5 \AA $\ $ in the source rest frame) emission line is present in the
optical spectrum. As a comparison 
in the RBS the fraction of emission line
AGN amongst the extragalactic sources is around 55\% (Schwope et al., 2000).
The few remaining non-emission line objects have
been classified as ``Normal Galaxies" or BL Lacs objects according
to the measured Calcium break discontinuity at the rest frame
wavelength of 4000 \AA $\ $ (see e.g. Landt et al., 2002). We stress
that some of the sources classified as ``Normal Galaxies"
could indeed host an AGN. As already discussed by Severgnini et
al. (2003) using X-ray and optical spectral data from this
project, the lack of significant emission lines in the optical
spectra can be explained by an adequate combination of the
absorption associated with the AGN and of the optical faintness of
the active nucleus with respect to the host galaxy. Furthermore
for some of the sources classified as ``Normal Galaxies" the
$H_{\alpha}$ line (in some case the only  spectroscopic evidence
of the presence of an AGN in the optical domain) is not sampled.
Although the presence of an AGN in the nucleus of some of these sources 
is highly probable (e.g. observed $L_x$ well in excess of $10^{42}$ \es)
we prefer to wait for a confirmation also from optical/infrared follow-ups; 
for the moment these objects are classified as ``Normal Galaxies".

To classify the emission line objects we have used the criteria
presented  for instance in Veron-Cetty \& Veron (2001) which are
based on the line width  and the line  flux ratios. Type 1 AGN
are those sources showing broad (FWHM $> 1000$ km s$^{-1}$)
permitted lines, while Type 2 AGN are those sources showing only
narrow lines (FWHM $< 1000$ km s$^{-1}$) and, when detected,
[OIII]$\lambda$5007/H$_{\beta} > $ 3.

A few sources show permitted lines with 1000 
km s$^{-1}$ $<$ FWHM $< 2000$ km s$^{-1}$ and
[OIII]$\lambda$5007/H$_{\beta}$ below 3. These sources are
probably narrow line Seyfert 1 candidates and, according to our
classification, have been included in the Type 1 AGN group.
For some sources classified as Type 2 AGN we have
indication of the presence of a broad component at the bottom  of
the narrow  H$_{\beta}$ and/or H$_{\alpha}$ lines.
These sources should be
properly classified as Seyfert 1.8 or Seyfert 1.9 objects;
for the purpose of the present paper these sources have been included in 
Type 2 AGN group. 
Finally for 26 objects a better S/N optical
spectrum and/or a more appropriate set-up for the spectroscopic
observations are needed to firmly classify them as Type 1 or
Type 2 AGN.  
These 26 sources have been marked in column 8 of Table 3 and 4.

\subsection {0.5-4.5 keV spectral properties}

A ``complete" spectral analysis for all the sources in the BSS and
HBSS samples (using data from the two EPIC MOSs and the EPIC pn)
is in progress; first results on selected sub-samples of sources have
been already discussed in Severgnini et al. (2003) and Caccianiga
et al. (2004). In the meantime, and in order to extract first order
X-ray spectral
information we present here a ``Hardness Ratio" analysis of the
single sources using only EPIC MOS2 data; this latter method is
equivalent to the ``color-color" analysis largely used at optical
wavelengths. The use here of the ``Hardness Ratio" analysis is
twofold. First of all it is much faster than a complete spectral
analysis with the combined use of three different instruments.
Second, a ``Hardness Ratio" is often the only X-ray spectral information
available for the faintest
sources in the XMM-Newton catalogue, and thus, a ``calibration" in the
parameter space is needed to select ``clean" and well-defined samples.
On the other hand, in Caccianiga et al. (2004) we have
already shown and discussed a tight correlation between X-ray absorption, as
deduced from a complete X-ray spectral analysis, and ``Hardness
Ratio" properties.

We have used the hardness ratios as defined from the XMM-Newton
pipeline processing\footnote {We have not used here the
``pipeline processing product" HR1 which is defined using the
corrected count rate in the (0.15$-$0.5) keV and in the (0.5$-$2)
keV  energy band since the measured count rate in the (0.15$-$0.5) keV band
is a strong function of the Galactic $N_{H}$
along the line of sight. Note that the effect on HR2 and HR3 due
to the different $N_{H_{Gal}}$ for the objects in the sample
(which ranges between $\sim 10^{20}$ to $10^{21}$ cm$^{-2}$) is
negligible. } :
~~~~~~\\
 \[
HR2={C(2-4.5\, {\rm keV})-C(0.5-2\, {\rm keV})\over C(2-4.5\, {\rm
keV})+C(0.5-2\, {\rm keV})}
\]
and
\[
HR3={C(4.5-7.5\, {\rm keV})-C(2-4.5\, {\rm keV})\over C(4.5-7.5\, {\rm
keV})+C(2-4.5\, {\rm keV})}
\]
~~~~~~\\
where C(0.5$-$2\, {\rm keV}), C(2$-$4.5\, {\rm keV}) and
C(4.5$-$7.5\, {\rm keV}) are the ``PSF and vignetting corrected"
count rates in the 0.5$-$2, 2$-$4.5 and 4.5$-$7.5 keV energy bands,
respectively.

In Figure 4 (panel a and panel b) we plot HR2 as a function of the 0.5--4.5
 keV
count rate for the extragalactic (and unidentified) sources of the
BSS sample. In particular, in panel a) we show the position of
the optically classified Type 1 and Type 2 AGN, while in panel b)
we have shown the unidentified sources, the
``Optically Normal Galaxies", the Clusters of galaxies and the
BL Lac objects. On the top, we have also reported the flux scale computed
assuming a conversion factor appropriate for $\alpha_E  \simeq 0.8$,
which is the ``mean" energy spectral index of the
``extragalactic" sample in the 0.5--4.5 keV energy band (see
below). 

No obvious trend in the source spectra as a function of the count rate is
measured. If we split the extragalactic BSS sample in two different bins of
count rate  (below or above a count rate of $1.41\times 10^{-2}$ cts/s in the
0.5-4.5 keV energy band) which includes a similar number of objects (163 and
170 sources, respectively) we found that the HR2 distributions of the two
sub-samples are consistent with being extracted from the same distribution 
with a probability of  61\% according to a KS test.  

\begin{table*}
\begin{center}
\caption{HR2 statistic for some relevant BSS and HBSS sub-sample(s).}
\begin{tabular}{lrrrrrr}
\hline
\hline
Sample               & Objects    & weighted             & unweighted         & median   & st.dev.   & $\sigma$ intrins.  \\
(1)                  & (2)        & (3)                  & (4)                & (5)      & (6)       & (7)                \\
\hline
BSS Sample:          &            &                      &                    &          &           &                \\
\ \ Extragalactic    &  333       & -0.51 $\pm$ 0.01     & -0.54 $\pm$ 0.01   & -0.57    &  0.20     & 0.16            \\
\ \ \ \ AGN1         &  180       & -0.54 $\pm$ 0.01     & -0.56 $\pm$ 0.01   & -0.58    &  0.15     & 0.12            \\
\ \ \ \ AGN2         &   26       & -0.32 $\pm$ 0.07     & -0.34 $\pm$ 0.07   & -0.49    &  0.35     & 0.32            \\
\ \ \ \ Unidentified &  111       & -0.52 $\pm$ 0.02     & -0.54 $\pm$ 0.02   & -0.56    &  0.16     & 0.12            \\
\ \ Stars            &   56       & -0.83 $\pm$ 0.01     & -0.88 $\pm$ 0.01   & -0.90    &  0.11     & 0.08            \\
HBSS Sample:         &            &                      &                    &          &           &                \\
\ \ Extragalactic    &   65       & -0.10 $\pm$ 0.05     & -0.27 $\pm$ 0.05   & -0.47    &  0.44     & 0.40            \\
\ \ \ \ AGN1         &   39       & -0.44 $\pm$ 0.03     & -0.49 $\pm$ 0.03   & -0.53    &  0.21     & 0.18            \\
\ \ \ \ AGN2         &   16       &  0.14 $\pm$ 0.11     &  0.05 $\pm$ 0.11   & -0.02    &  0.45     & 0.44            \\
\hline \hline
\end{tabular}
\end{center}
Columns are as follows:
(1) Sample;
(2) Number of sources;
(3) HR2 weighted average;
(4) HR2 unweighted average;
(5) HR2 median value;
(6) observed standard deviation of the HR2 distribution;
(7) intrinsic dispersion of the HR2 distribution computed
using the maximum likelihood method as described in
Maccacaro et al, 1988.
\end{table*}

The weighted average of the extragalactic population is HR2$=-0.51\pm 0.01$
corresponding to $\alpha_E = 0.74\pm 0.03$ (the relation between the spectral
index and HR2 has been tabulated for few representative spectral indices in
Table 1). For comparison the weighted averages of the sources classified as Type
1 AGN and unidentified objects are HR2$=-0.54\pm 0.01$ and HR2$=-0.52\pm 0.02$,
respectively. The HR2 distribution of the sources identified as Type 1 AGN
appears to be ``narrow" with $\sim 90$\% of the sources inside the HR2 range
$-0.75$ to $-0.35$. On the contrary the Type 2 AGN are characterized by a
broader distribution with $\sim 42$\% of the objects having an observed energy
spectral index apparently flatter than that of the cosmic X-ray background
($\alpha_E = 0.4$ corresponding to HR2$=-0.38$). For comparison only $\sim 10$\% 
of the unidentified X-ray sources or of the X-ray sources identified as
broad line AGN seem to have spectra apparently flatter than $\alpha_E = 0.4$.

Please note that two broad line AGN, which are clearly  separated from the 
majority of the other broad line AGN, seem to be  characterized by an extremely
hard spectrum ($\alpha_E \sim -1$). These two objects, belonging both  to the
BSS and HBSS samples,  are XBS J091828.4+513931 (HR2=0.31) and XBS
J143835.1+642928 (HR2=0.18). For the first object the optical spectra show
broad $H_{\beta}$ and  $H_{\alpha}$ emission lines without any obvious sign of
peculiarity, while in the case of XBS J143835.1+642928 the optical spectrum in hand
is very noisy and therefore the optical classification is at the moment
tentative.  The X-ray spectra of both sources are described by an  absorbed
power-law model having an intrinsic $N_H$ in excess of $10^{22}$ cm$^{-2}$.  At
the moment these two sources are the only broad line AGN displaying intrinsic
absorption (as derived from a complete X-ray spectral analysis) above $10^{22}$
cm$^{-2}$ but the completion of the X-ray spectral work for the total sample
is  needed to evaluate correctly the fraction of X-ray absorbed  broad line
AGN in this survey
\footnote{We note that in the HBS28 sample discussed in Caccianiga et al.,
2004  none of the 19 Type 1 AGN is absorbed with $N_H$  in excess to $10^{22}$ 
cm$^{-2}$, implying that the fraction of X-ray absorbed Type 1 AGN is less 
than 10\% of the total Type 1 AGN population. This result is also confirmed 
by the present HBSS sample since at most 4 (out of 39) broad line AGN are
characterized by HR2 values typical of absorbed ($N_H > 10^{22}$ cm$^{-2}$) 
AGN. See also section 3.4.}. 

It is now worth comparing the HR2 properties of the extragalactic BSS and HBSS
sources; in figure 4 (panel c and d) we plot HR2 as a function of the 4.5--7.5
keV count rate for the extragalactic (and unidentified) sources of the HBSS
sample. As a class, the Type 1 AGN in the HBSS sample seem to have the same
HR2, and thus 0.5--4.5 keV spectral properties, of the Type 1 AGN in the BSS;
about 87\% of them reside in the HR2 range between $-0.75$ and $-0.35$, with a
median HR2 value of $-0.53$. On the contrary the Type 2 AGN in the HBSS seem to
be characterized by  more extreme spectral properties if compared with the Type
2 AGN in the BSS sample; $\sim 81$\% of them seem to have an energy spectral
index flatter than 0.4 and $\sim 2/3$ seem to have inverted spectra
($\alpha_E < 0$). It is worth noting that 2 of the 3 type 2 AGNs having HR2
around $\sim -0.55$ in the HBSS sample are Seyfert 1.9 galaxies.

The HR2 statistic for some relevant BSS and  HBSS sub-sample(s) have been
summarized in Table 7.

We note that in the case of the extragalactic population in the BSS sample the
use of the weighted average, the unweighted average or the median value of HR2 
give  consistent results on the underlying spectral index ($\alpha_E \sim
0.74$, $\alpha_E \sim 0.83$ and $\alpha_E \sim 0.91$, respectively). On the
contrary in the case of the extragalactic population in the HBSS sample  the
use of the  weighted average, the unweighted average or the median value  of
HR2 gives  completely different results ($\alpha_E \sim -0.27$, $\alpha_E \sim
0.11$ and $\alpha_E \sim 0.64$,  respectively). 
Since the HR2 distribution of
the extragalactic population in the HBSS sample is significantly 
different  from a Gaussian we
prefer to use as   ``average" spectral index of this sample that related to the
median value ($\alpha_E \sim 0.64$).

A similar result has been pointed out by Nandra et al., 2003 studying  the
spectral properties of the sources in the SHEEP (5-10 keV) survey.  These
authors prefer the use of the unweighted average over the  weighted average
(they did not consider median values) and reach the result that $<\alpha_E>\sim
0$ and the conclusion that the 5--10 keV surveys are sampling a completely
different population compared with the 2-10 keV surveys. 

On the contrary we find strong evidence that the 0.5-4.5  spectral properties
of the class of broad line AGN in the BSS and in the  HBSS are very similar.
Moreover, since the majority of the objects in the BSS and HBSS samples are in
common we do not find compelling evidence that the surveys in the two bands
are selecting completely different populations. The HBSS survey  is simply more
efficient than the BSS survey in selecting the hard part of the intrinsic source
spectral distribution. 

The eleven objects belonging only to the HBSS sample (enclosed inside empty squares
in figure 4, 5 and 6 below) are amongst the  hardest
X-ray sources in the sample;  all but one (XBS J140113.4+024016, still optically
unidentified) of them seem to be characterized by an
apparently inverted spectrum ($\alpha_E < 0.0$).  Among these sources there
are 5 Type 2 AGN, one optically normal galaxy,  one BLLac object and 4
unidentified objects.

\begin{figure*}
\begin{center}
\begin{tabular}{cc}
\includegraphics[width=0.50\textwidth]{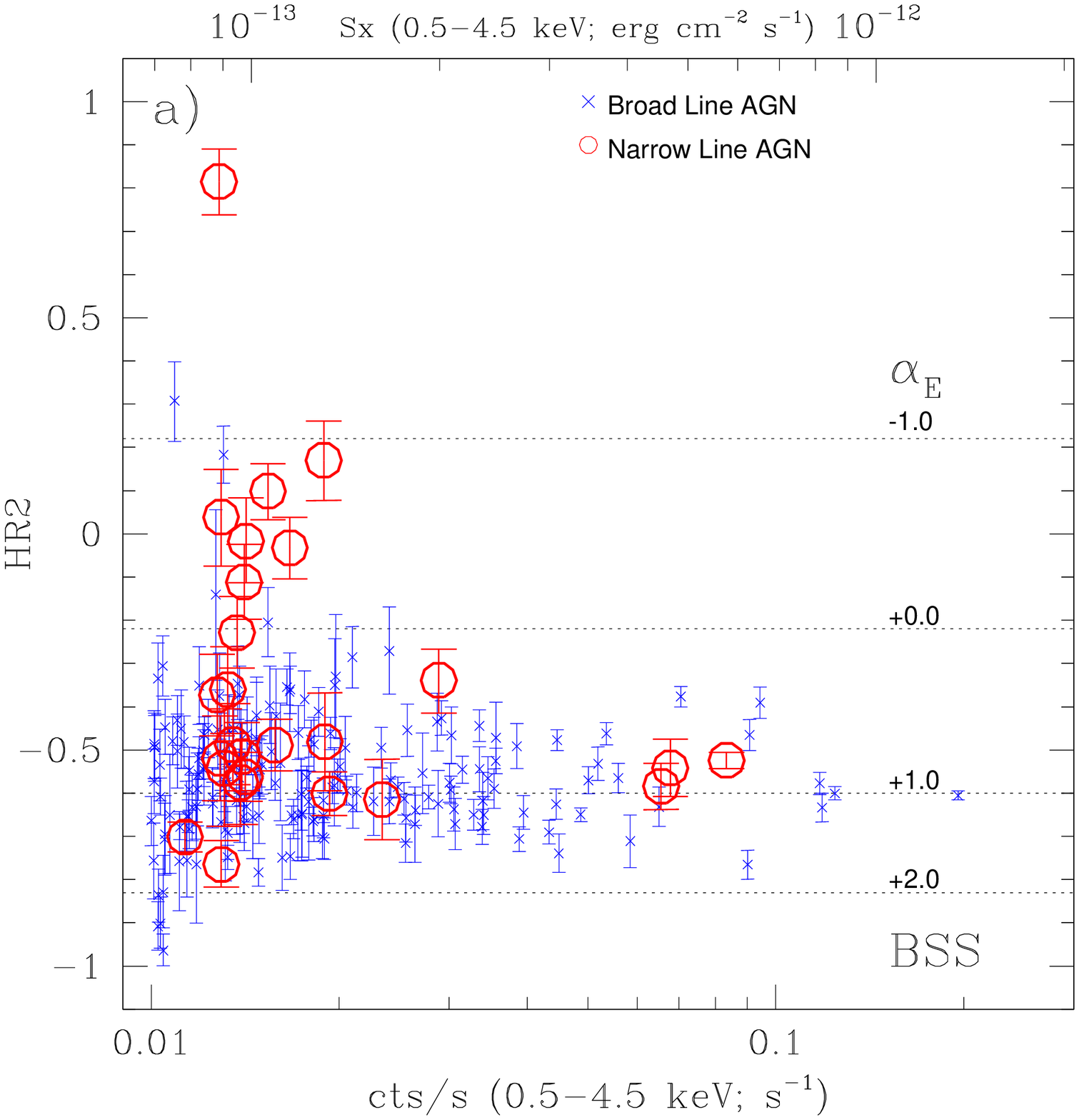}&\includegraphics[width=0.50\textwidth]{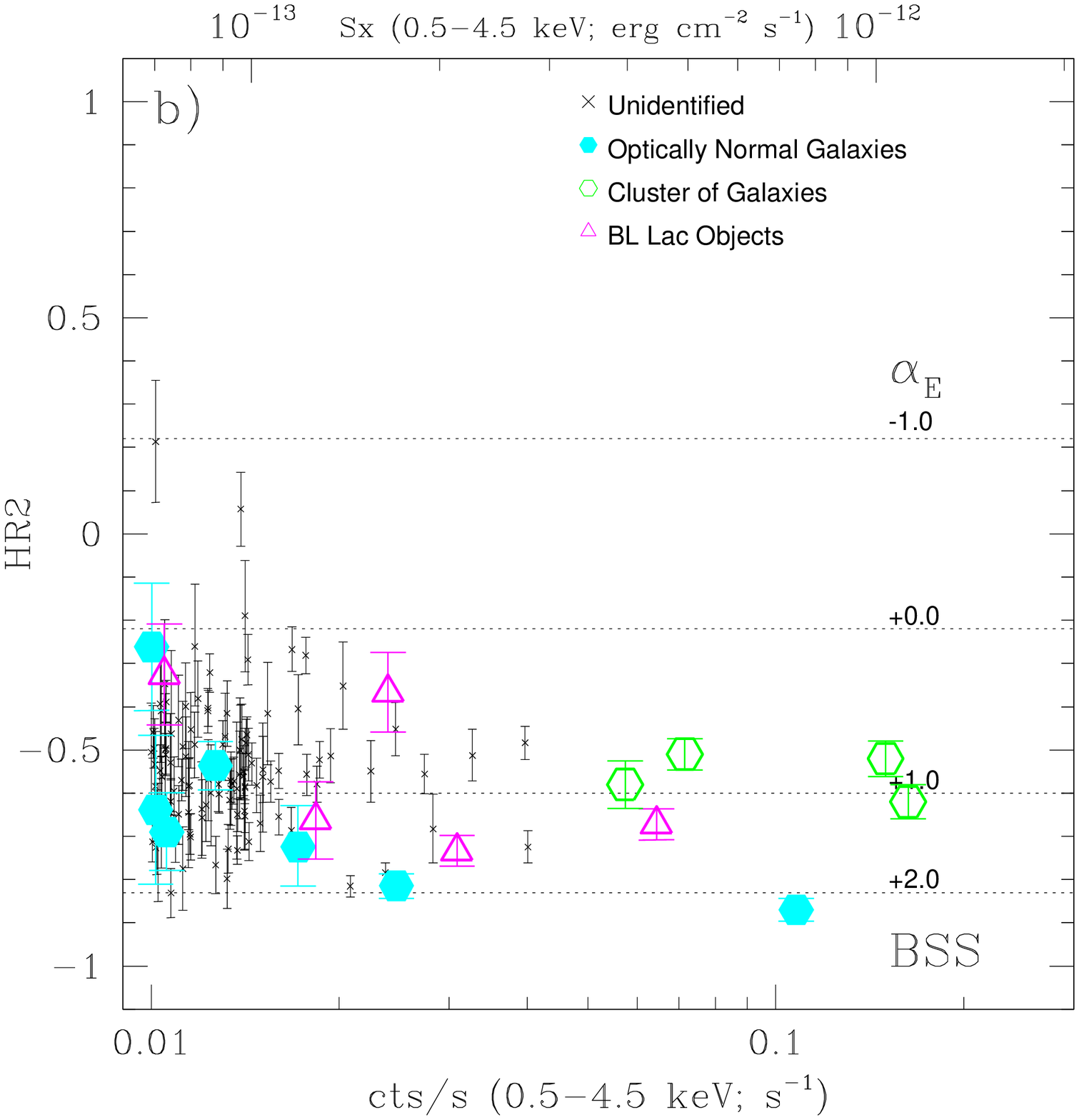}\\
\includegraphics[width=0.50\textwidth]{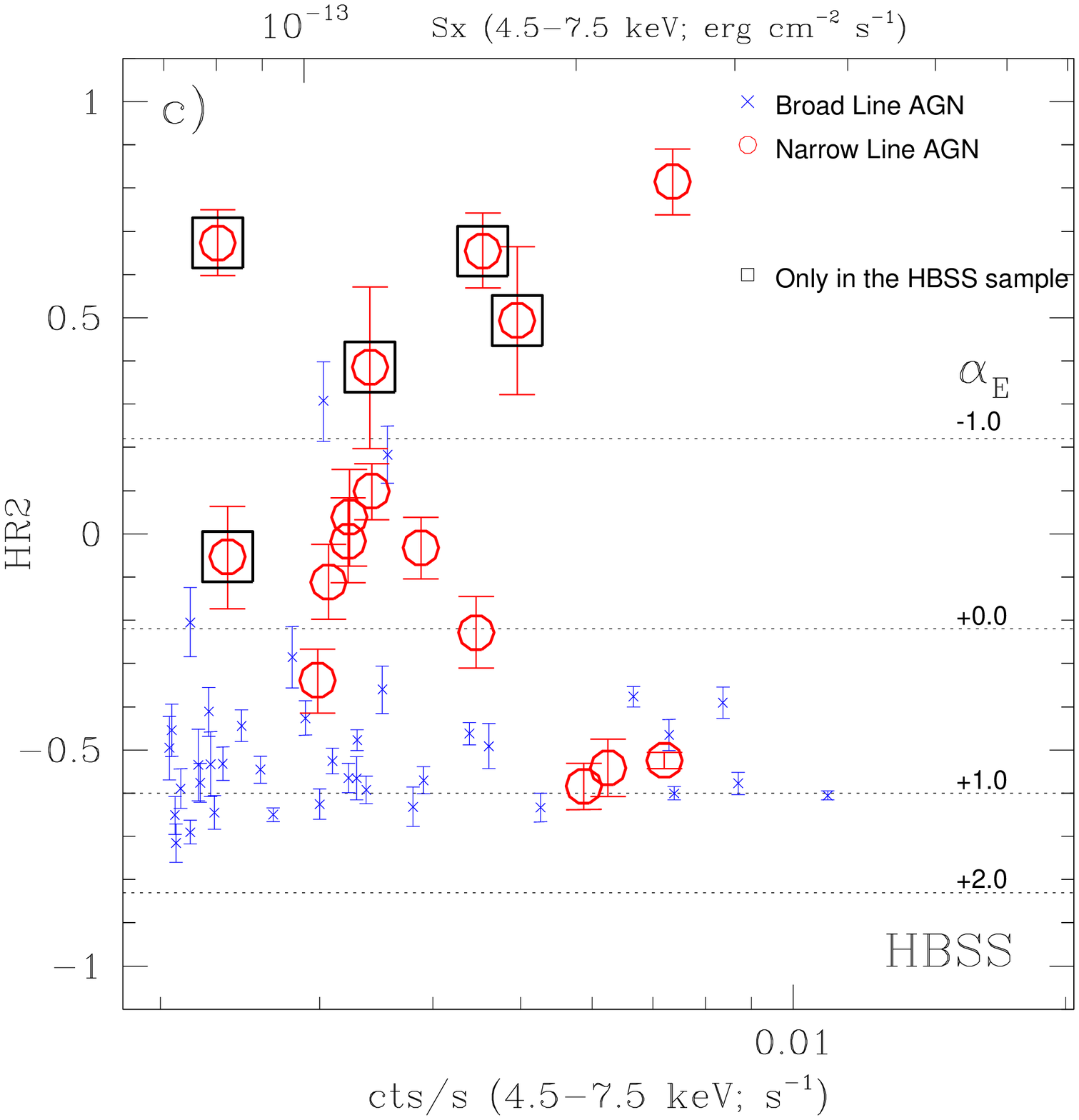}&\includegraphics[width=0.50\textwidth]{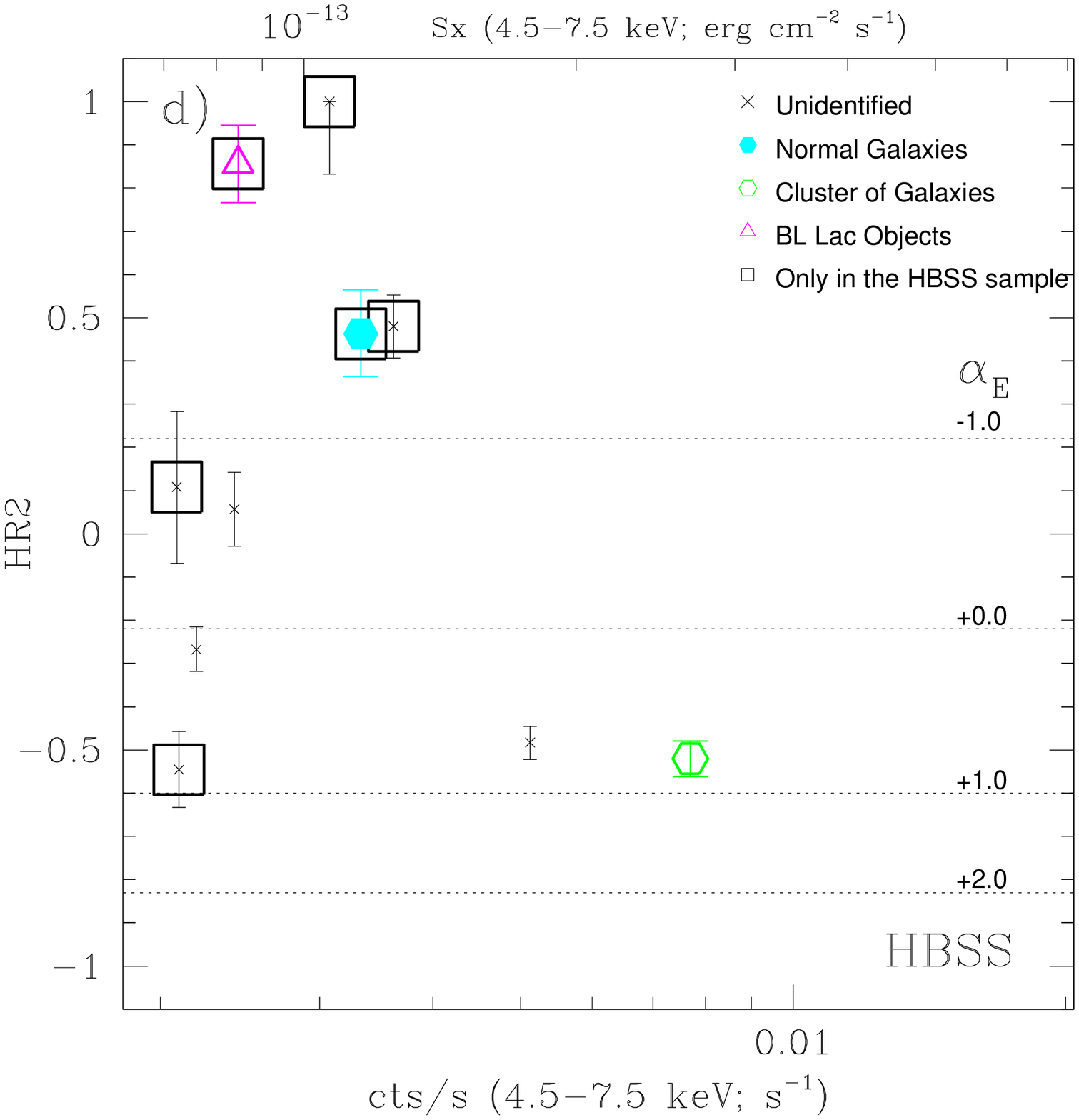}\\
\end{tabular}
\caption{ HR2 vs. EPIC MOS2 count rate in the selection band for
the sources in the BSS (panel a and panel b) and  HBSS sample
(panel c and panel d). We have also reported  the HR2 expected
from a unabsorbed power-law model with $\alpha_E$ ranging from
-1 to 2 ($S_E \propto E^{-\alpha_E}$). The flux scale on the top
has been computed assuming a conversion factor appropriate for
$\alpha_E \sim 0.8$ (BSS sample) or $\alpha_E \sim 0.7$ (HBSS
sample). We have used different symbols to mark the different
kinds of objects. The eleven sources belonging only to the HBSS sample 
are enclosed inside empty squares in panels c and d. 
}
\end{center}
\end{figure*}

\subsection {Broad-band X-ray Spectral Properties}

 Combining the information on HR2 and HR3 we can now investigate
in more detail the broad band  spectral properties of the
sample(s) as well as the selection function(s) of the BSS and
HBSS.

\begin{figure*}
\begin{center}
\begin{tabular}{cc}
\includegraphics[width=0.50\textwidth]{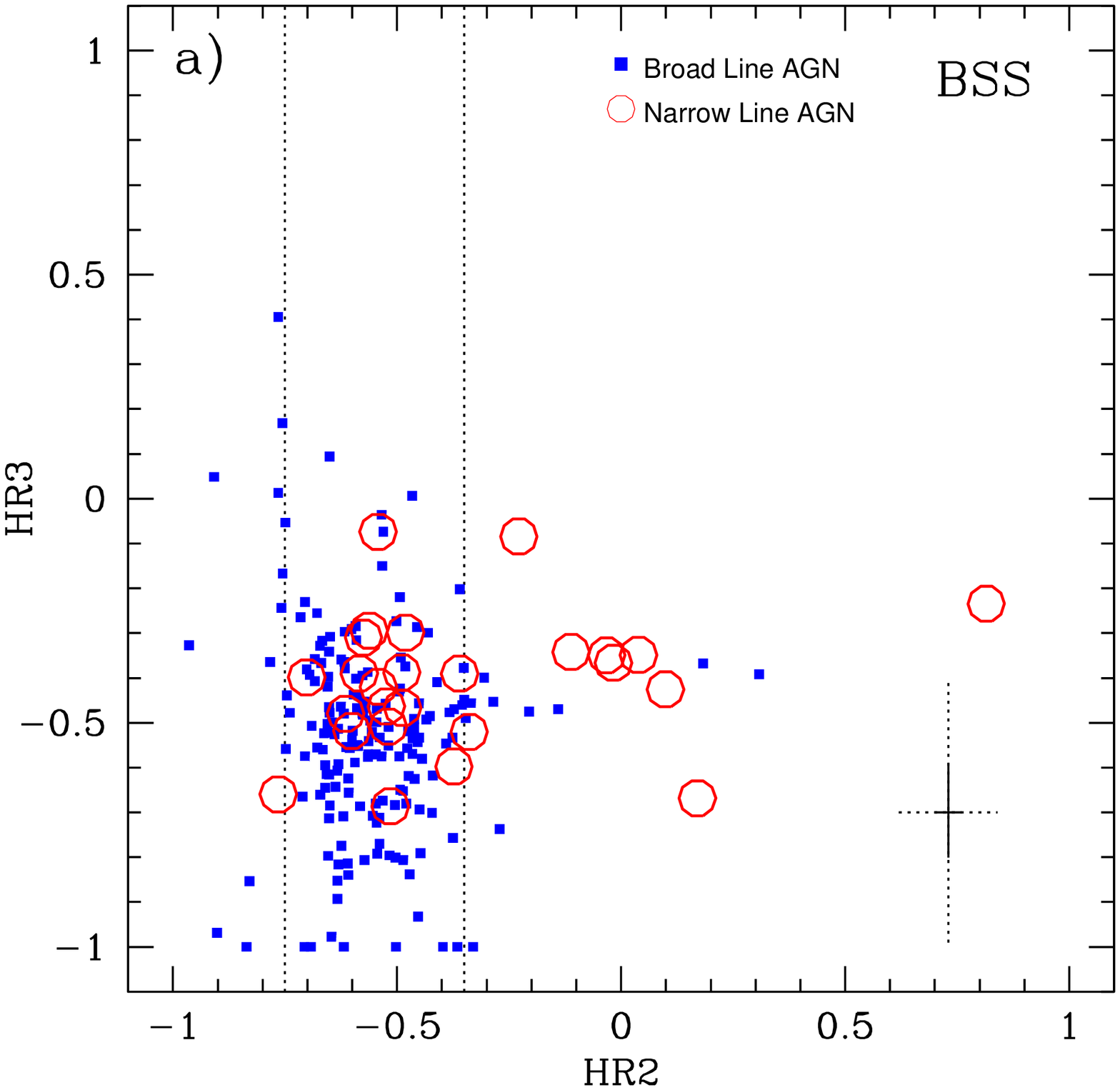}&\includegraphics[width=0.50\textwidth]{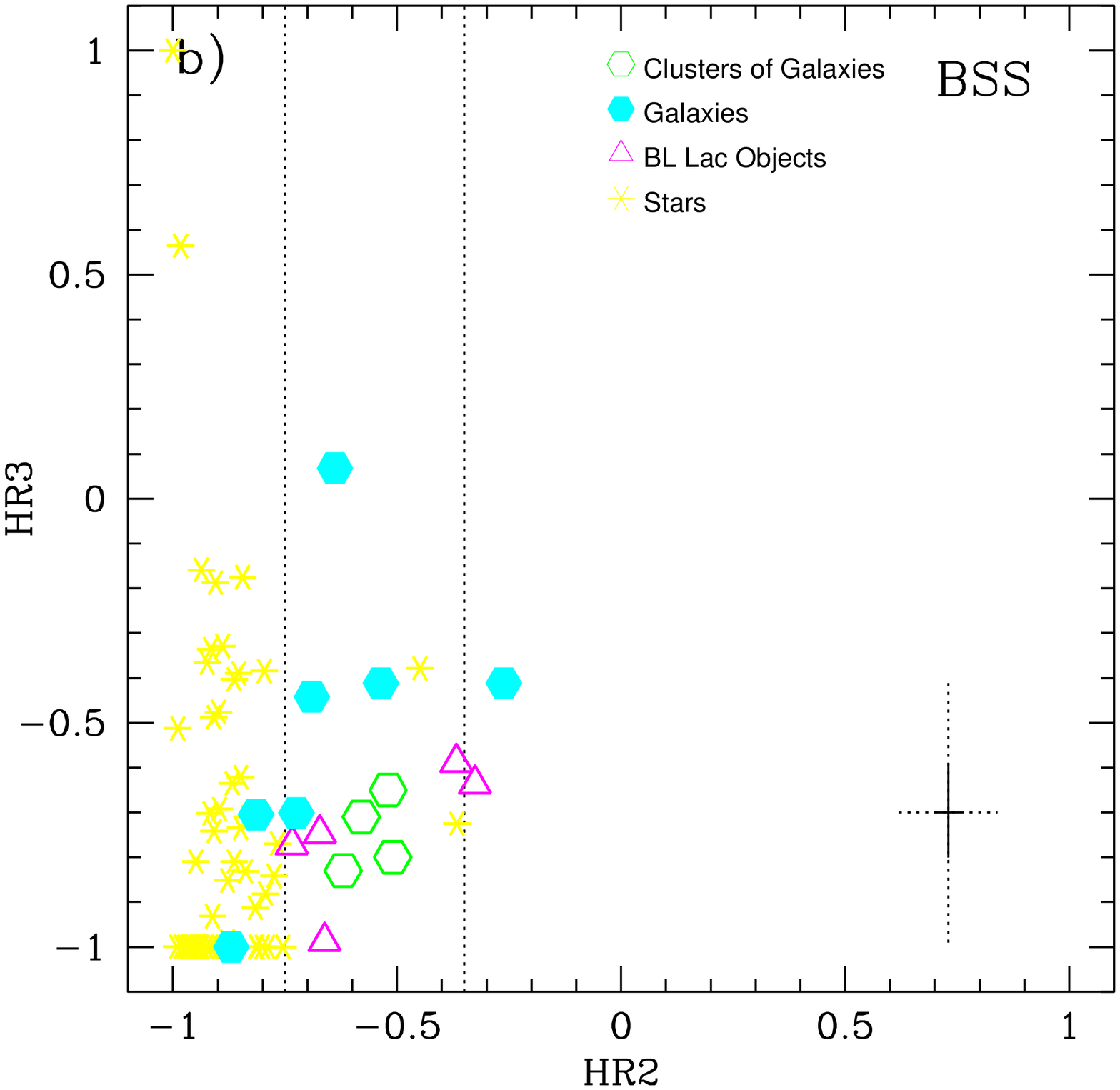}\\
\includegraphics[width=0.50\textwidth]{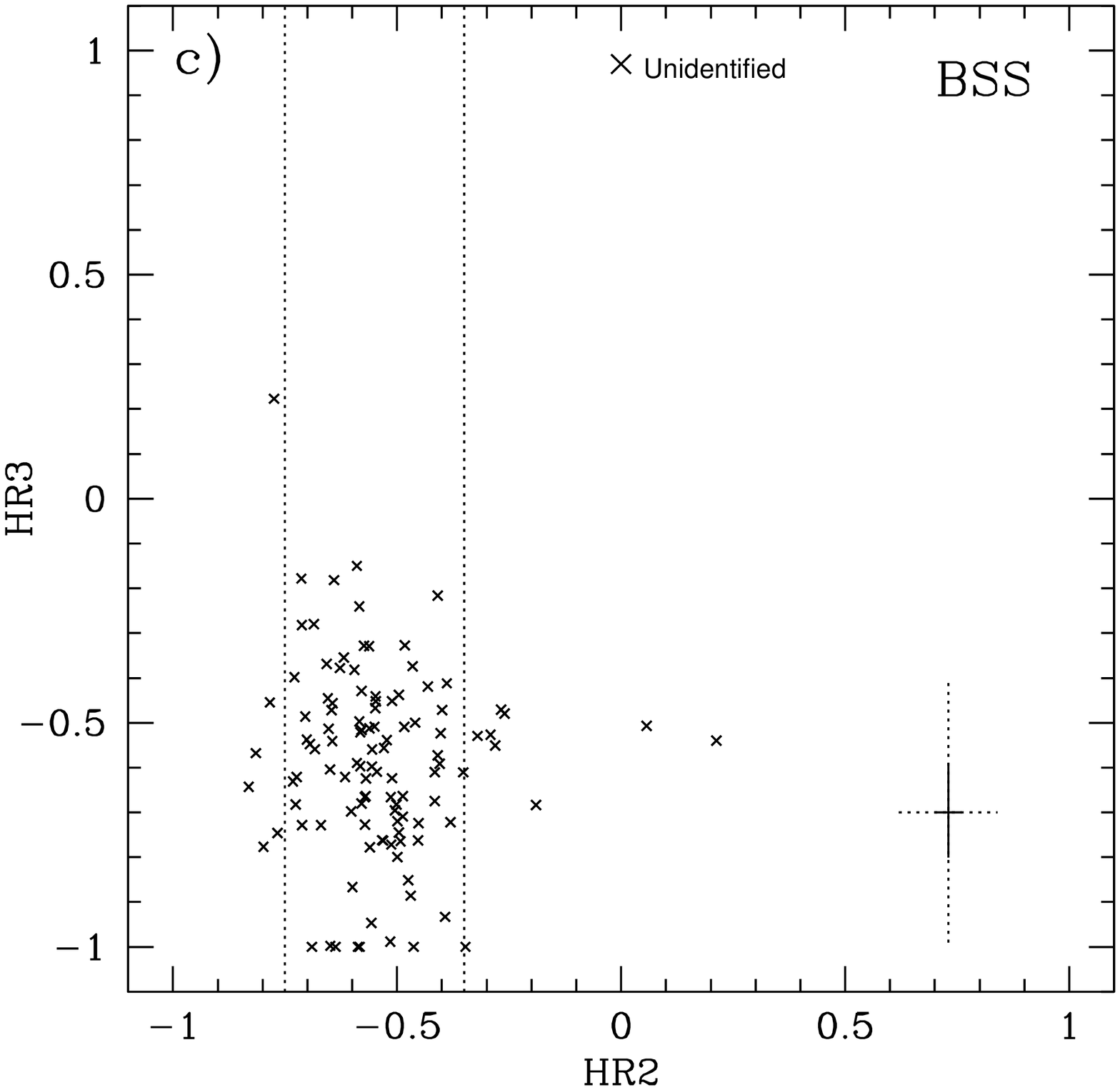}&\includegraphics[width=0.50\textwidth]{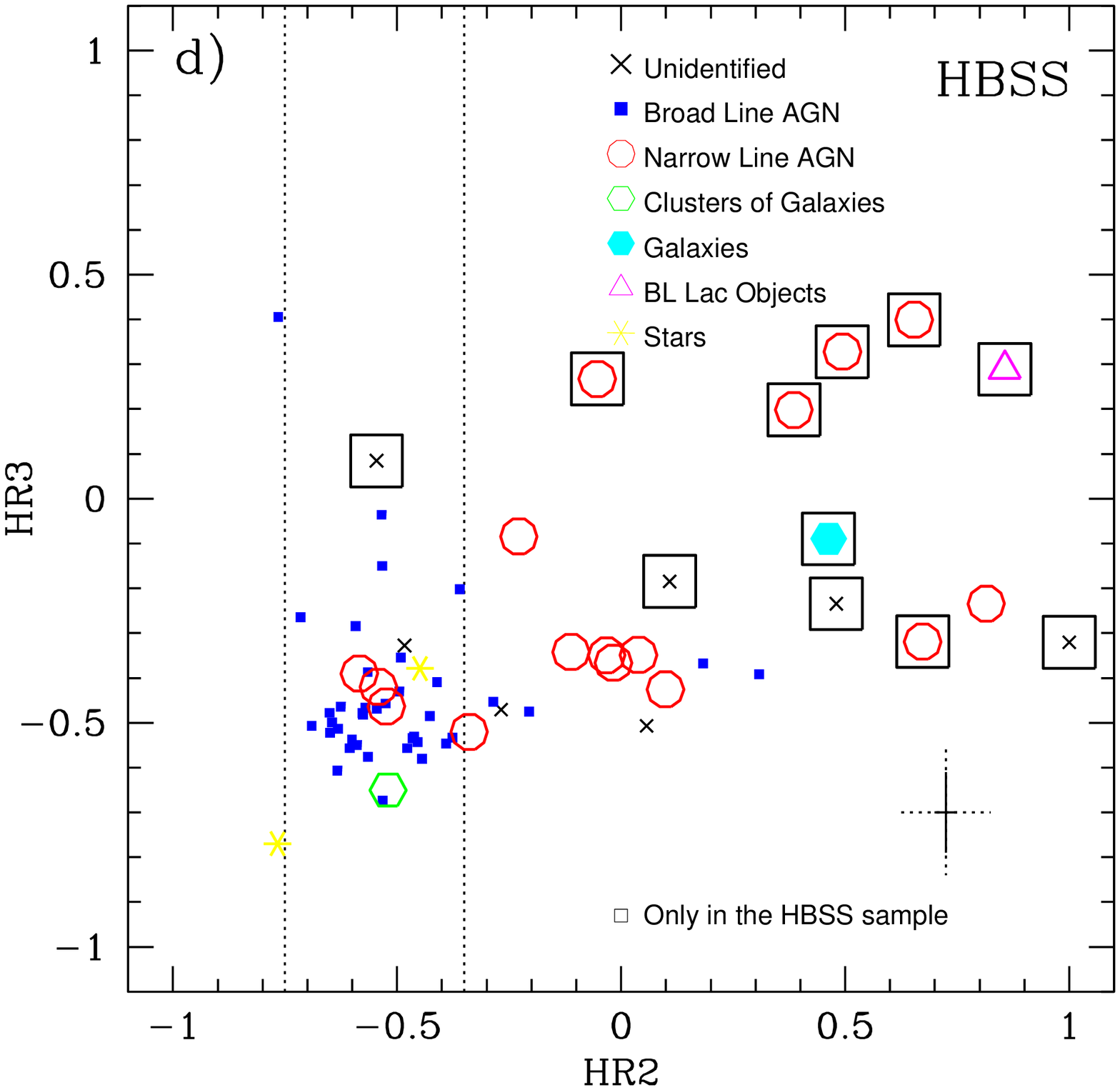}\\
\end{tabular}
\caption{ HR2 vs. HR3 for the sources belonging to the BSS
sample (panel a, b and c) and to the HBSS sample (panel d). The
dotted lines at constant HR2 correspond to the locus enclosing
$\sim 90$\% of the Type 1 AGN in the BSS sample; these lines have
been reported in all panels to assist with the comparison(s).
We have used different symbols to mark the
spectroscopically identified and unidentified objects in the two
samples. The eleven sources belonging only to the HBSS sample 
are enclosed inside empty squares in panel d. 
In the lower right corner of each panel we have also
reported the median error on HR2 and HR3 for the total BSS and
HBSS sample (solid line) and the 90\% percentile on these errors
(dotted line).}
\end{center}
\end{figure*}

The comparison is made in figure 5 where we show the position of
the BSS (panels a, b and c) and HBSS (panel d) sources in the
HR2-HR3 plane. We have used different symbols and panels to mark
the spectroscopically identified and unidentified objects.
Useful information can be extracted by
cross-comparing the position of the different optical types of
sources as well as by comparing  the position of the sources in
the BSS and HBSS samples.

The ``bulk" of the sources optically identified as broad line AGNs
are strongly clustered in the region between HR2=$-0.75$ and
HR2=$-0.35$; this is true both for the broad line AGNs belonging
to the BSS sample (by definition) and for those belonging to the
HBSS sample (only 5 of 39 HBSS broad line AGN are outside these
limits). The spread on HR3 is much larger than the spread on HR2
but note that HR3 is much noisier than HR2 since many of the
sources are detected with poor statistics (or even undetected) in
the 4.5--7.5 keV energy band.

All but 2 ($\sim 96\%$) of the sources classified as
stars\footnote{The two sources that are outside the ``stellar" HR2
locus are XBS J014100.6$-$675328 and XBS J215323.7+173018. 
The first object, XBS J014100.6$-$675328, has been already 
discussed in footnote 9.
The XMM spectrum, discussed in Caccianiga et
al., 2004, is well fitted by a unabsorbed power-law model with
$\Gamma \sim 1.53$ plus a thermal component having a temperature
$kT\sim 55$ eV and an emission line at $E\sim 6.7$ keV. The second
object (XBS J215323.7+173018) is a star having an XMM spectrum
described by a thermal component with $kT\sim 1.6$
keV and a possible hard tail.} 
in the BSS sample have an HR2 less
than $-0.75$. If we assume a simple Raymond-Smith thermal model,
HR2$\leq -$0.75 corresponds to temperatures below $\sim 1.5$ keV,
in very good agreement with the identification as coronal emitting
stars.

In Caccianiga et al., 2004, using a ``pilot" 4.5$-$7.5 keV sample composed of
28 X-ray sources (26 of which in common with the current version of the HBSS
sample reported in Table 4), we have already discussed the correlation between
X-ray absorption, as deduced from a complete X-ray spectral analysis, and
``Hardness Ratio" properties. In particular we have found that
a) at the sampled fluxes, the 4.5$-$7.5 keV selection is picking up AGN having 
an intrinsic $N_H$ up to few times $10^{23}$ cm$^{-2}$; 
b) all the AGN having HR2$>-0.35$ are X-ray absorbed with $N_H$ ranging from
few times 10$^{21}$ up to few times 10$^{23}$ cm$^{-2}$. 
Assuming that this result is valid also for the sources presented here  we
can conjecture that all the sources having HR2 greater than $-0.35$ are 
absorbed AGN; in this part of the diagram, besides narrow line AGN, we also
note a few broad line AGN and 2 sources 
(one in the BSS sample and one in the HBSS sample) 
optically identified as normal galaxies. However also in these
sources their point-like X-ray emission, their X-ray spectra 
(a preliminary spectral analysis shows that they are described by an
absorbed power-law model having $N_H >\sim  10^{22}$ cm$^{-2}$) and their
intrinsic luminosity (in excess of $\sim 7 \times 10^{43}$ \es in the 2-10 keV
energy range) strongly suggest the presence of an AGN.  The existence of
relatively luminous X-ray sources, optically identified with  ``normal
galaxies", has been reported since the {\it Einstein}  Observatory era in the
early eighties (Elvis et al., 1981);  this kind of sources were called in a
variety of names such as  optically dull galaxies (Elvis et al., 1981), passive
galaxies  (Griffiths et al., 1995) and X-ray bright optically normal galaxies 
(XBONG, Comastri et al., 2002).  We have already discussed in Severgnini et al.
(2003) that detailed and specific optical-infrared follow-ups or higher-quality
optical spectra  are needed to unveil the AGN also in the optical domain.
An advection-dominated accretion flow model has been recently used  
by Yuan and Narayan (2004) to explain their broad band properties.

Contrary to broad line AGNs and stars, narrow line (Type 2) AGNs seem to be
distributed over a larger area in the HR2$-$HR3 plane with a well visible
difference in the source position between the Type 2 AGNs in BSS and those in
the HBSS sample.  Although many of them have the hardest spectra amongst the
identified objects, highly suggestive of intrinsic absorption (see also figure
4), a new fact  which seems to emerge from this investigation is the large
number of narrow line AGN in the BSS sample occupying the locus typical of
X-ray unabsorbed broad line AGN.  Taking into account the still incomplete
spectroscopic identification work  and that some sources need a better quality
optical spectrum we estimate  that the relative fraction of these objects over
the entire Type 2 AGN  population may range between 50 and 75\% in the case of
the BSS sample and around  20\% in the case of the HBSS sample 
\footnote {This latter estimate is consistent with the fact that in the  
HBS28 sample
discussed in Caccianiga et al., 2004 there is only one Type 2 AGN  with HR2
between $-0.75$ and $-0.35$ out of 8 Type 2 objects.}.
It is also worth noting that  2 out of 4 of the
Type 2 AGN belonging to the ($2-10$ keV) HELLAS2XMM survey and having a good
X-ray statistic (sample S1 in Perola et al.,  2004) are characterized by an
``observed" intrinsic $N_H$ well below  $10^{22}$ cm$^{-2}$. 

To our knowledge two kinds of narrow line AGN could populate this zone, and
thus could have X-ray spectral properties similar to those expected from
unabsorbed AGN: a) ``Compton thick" absorbed AGN (see e.g. the results
presented in Della Ceca et al., 1999 from ASCA data); b) objects similar to the
class of unabsorbed Seyfert 2 discussed in Pappa et al. (2001); Panessa and
Bassani (2002) and in Barcons, Carrera and Ceballos (2003).  Also the
variability could play some role if the nucleus was bright at the  time of the
XMM-Newton observation but was turned off at the  time of the optical
spectroscopy. A detailed and exhaustive analysis of these sources is beyond the
scope of the  present paper.  A deeper investigation of their optical  (e.g.
finer optical classification between Seyfert 1.8, Seyfert 1.9 and Seyfert 2, 
analysis of the O[III] to 2-10 keV flux ratio) 
and X-ray (e.g. presence of Fe $K_\alpha$ emission lines to  evaluate the
Compton-thickness of the source)  properties, as well as an assessment of the
role played by selection effects, is in progress and will be presented
elsewhere.

\subsection {X-ray to optical flux ratio}

A useful parameter to discriminate between different classes of
X-ray sources is the X-ray to optical flux ratio 
(X/O flux ratio hereafter; see Maccacaro et al., 1988).

Previous investigations (see Fiore et al., 2003 and references therein)
have shown that 
standard X-ray selected AGN (both type 1 and type 2) have typical X/O flux ratios
in the range  between 0.1 and 10 (for comparison  standard optically selected
QSOs and  Seyfert 1 galaxies have  X/O flux ratio  $\sim 1$).  X/O flux ratios 
below 0.1 are
typical of coronal emitting 
stars, normal galaxies (both early type  and starforming) and nearby
heavily absorbed (Compton thick) AGN. Finally at high  X/O flux ratios (well above
10) we can find broad and narrow line AGN as well as 
high-z high-luminosity  obscured AGN (type 2 QSOs), high-z
clusters of galaxies and extreme BL Lac objects.

In this paper we have defined the X-ray to optical
flux ratio  using the observed X-ray flux in
the 0.5--4.5 keV energy range and the optical red-band
flux (see Fukugita et al., 1995 for the appropriate conversion factors).

For the large majority (73\%) of the objects we have
found and used APM red magnitudes. For 19\% of the objects red
magnitudes have been found from the literature, measured by us
during spectroscopic runs or estimated using magnitudes measured
in other optical bands. Finally for the unidentified objects with
optical magnitudes below the POSS II limit (30 sources) we have
used an upper limit of $mag_R\simeq 21.0$. We note here that, for a
fixed X-ray flux, an error of $\sim 1$ magnitude corresponds to an
uncertainty of $\sim 60$\% on the X/O flux ratio; this uncertainty does
not affect any of the general conclusions we discuss below.

In figure 6 we have plotted the X/O flux ratio versus the  HR2 value for each
source (BSS sample in panels a and b; HBSS sample in panel c). The boxes
defined by the dot dashed lines (dashed line)  indicate the locus of 
coronal emitting stars
(broad line AGN) in the BSS sample and have been reported in all panels to
assist with the comparisons.  

As is clearly visible in figure 6 the bulk of coronal emitting stars is well
separated from the bulk of extragalactic sources; some of the AGN (both broad
and narrow line)  have X/O flux ratio typical of stars but HR2 values typical of
AGN; similarly some AGN with an HR2 typical of stars can be distinguished from
stars thanks to their X/O flux ratio.  Therefore the combined use of X/O flux ratio
and HR2 allows us to distinguish almost unequivocally galactic sources from the
extragalactic ones.

Around 10\% of the extragalactic population have a X/O flux ratio greater than 10. 
If we consider the 2--10 keV fluxes instead of the 0.5--4.5 keV fluxes this 
fraction increases to $\sim 15$\%, in good agreement with the results obtained 
by Fiore et al., 2003 at fainter fluxes.
Amongst the sources identified so far at X/O flux ratio $>$ 10 there
are some broad line and a few narrow line AGN. These results are consistent
with those reported in Fiore et al., 2003 who also  discuss the observational
evidence that at X/O flux ratio greater than 10 is where to look for type 2 QSOs. 
Since the large
majority of the X-ray sources at X/O flux ratio $>$ 10 presented here  is  still
unidentified we can not comment further on this; we can only note that many of
the unidentified objects with high or very high X/O flux ratio seem to have  rather
standard  hardness ratios.

The opposite side of the extragalactic X/O distribution  (X/O flux ratio
$<0.1$)  is populated by  optically normal galaxies, Type 2 AGNs and a few
broad line AGN. As discussed above this part of the X/O  distribution could be
also populated  by nearby Compton thick AGN;  in this respect it is worth 
noting  that a few of the type 2 AGN in the BSS sample populating the same HR2 
region of the broad line AGN population have X/O flux ratios $<0.1$.   In the
HBS sample there are also a few Type 2 AGN having apparently hard spectra  but
low  X/O flux ratios ($<0.1$).

\begin{figure*}
\begin{center}
\begin{tabular}{cc}
\includegraphics[width=0.50\textwidth]{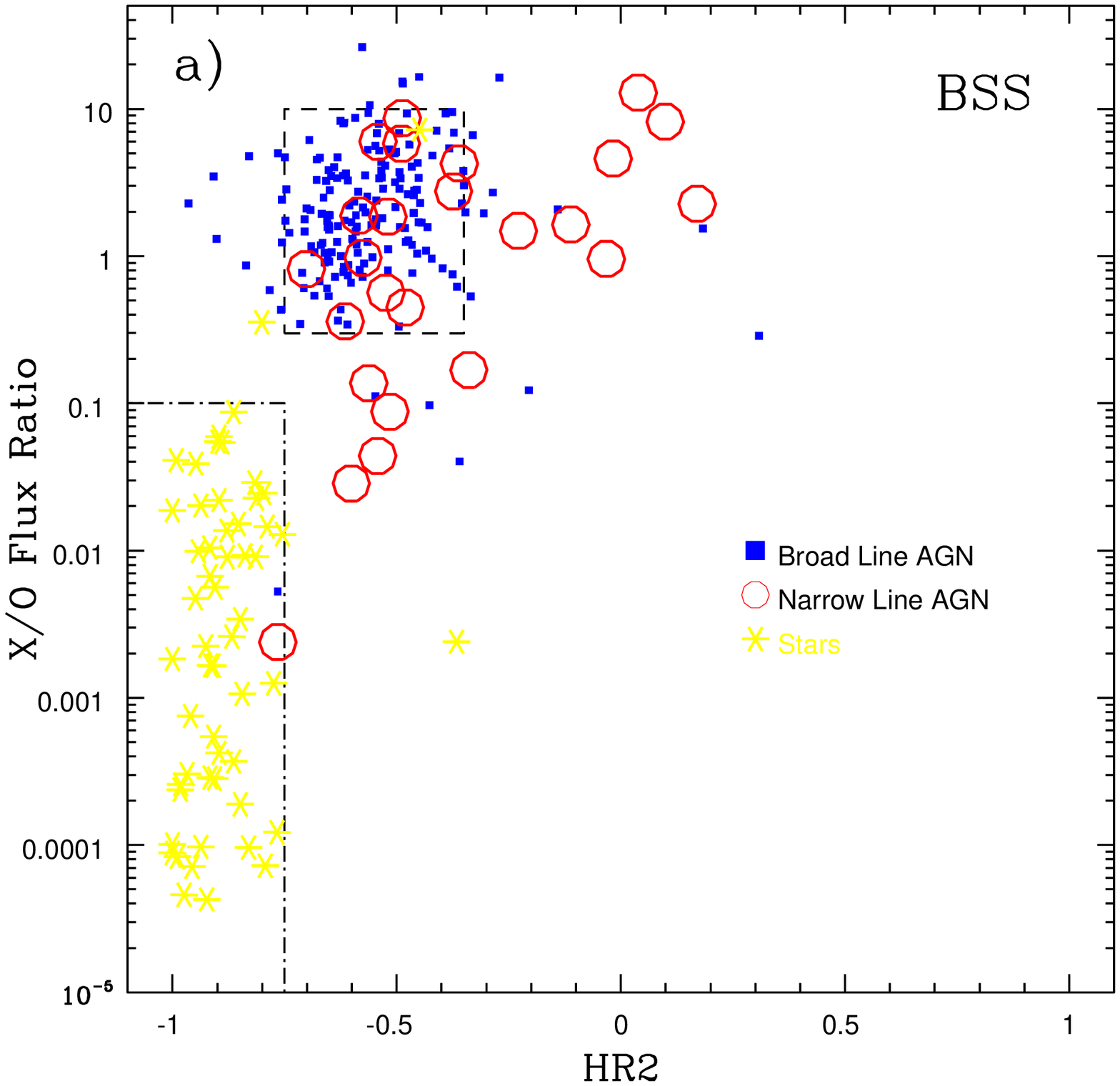}&\includegraphics[width=0.50\textwidth]{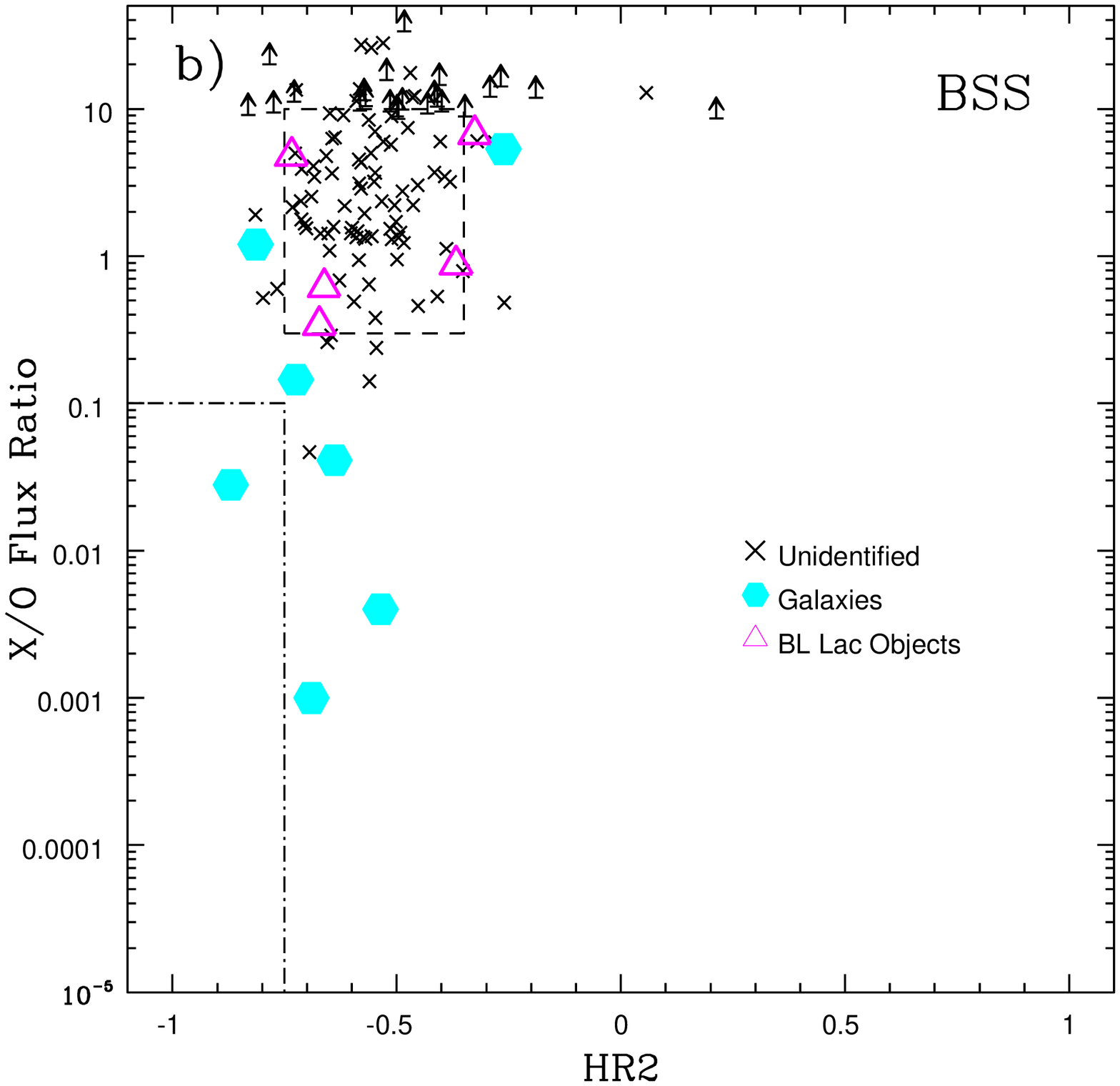}\\
\includegraphics[width=0.50\textwidth]{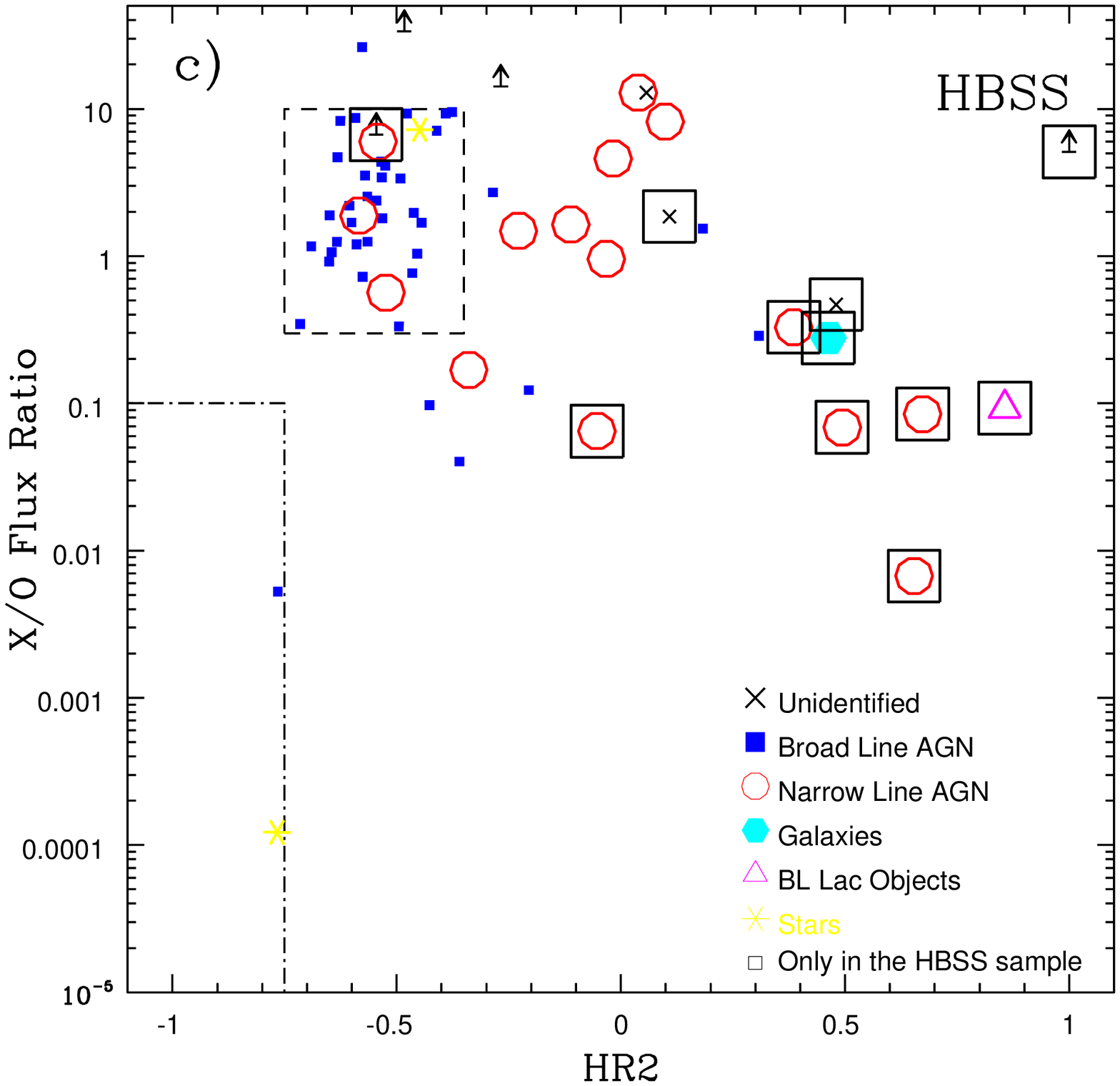}&
\begin{minipage}{9cm}
\vspace*{-7cm} \caption{ HR2 vs. X-ray (0.5-4.5 keV) to optical
flux ratio for the sources belonging to the BSS sample (panels
a and b) and to the HBSS sample (panel c). The X-ray
(0.5-4.5 keV) to optical flux ratio for each source has been
computed as discussed in section 3.5. We have used different
symbols to mark the spectroscopically identified and unidentified
objects in the two samples. The box defined by the dot dashed
lines (dashed line) enclose about 95\% (85\%) of the sources
optically identified as coronal emitting stars (broad line AGNs) in the BSS
sample; these boxes have been reported in all panels to assist with
the comparison(s) discussed in the text.
The eleven sources belonging only to the HBSS sample 
are enclosed inside empty squares in panel c.}
\end{minipage}\\
\end{tabular}
\end{center}
\end{figure*}

\subsection {The number densities of optically broad and narrow line AGN}

The measured X-ray Log(N$>$S)-LogS of broad and narrow line AGN is a
fundamental observable for cosmological investigations and, in particular,
provides very strong constraints on the CXB synthesis models and on the history
of the accretion  in the Universe.

As discussed above the spectroscopic identification rate of the HBSS sample is
$\sim 90$\%, allowing us to investigate for the first time the X-ray 
Log(N$>$S)-LogS of optically broad and narrow line AGN in the same sample. In order to
deal with the 7 unidentified HBSS sources we have made the reasonable
assumption that the 5 unidentified sources having HR2$>-0.35$ are Type 2 AGN, 
while the 2 unidentified  sources having HR2 in the range from $-0.75$ to
$-0.35$ are Type 1 AGN (see section 3.3 for details).

The 4.5--7.5 keV Log(N$>$S)-LogS of optically broad and narrow line AGN are reported in
figure 7 (AGN1: open circles; AGN2: filled circles). A conversion factor
appropriate for a  power-law spectral model with $\alpha_E \simeq 0.7$ has been
used to obtain the fluxes. Both
LogN($>$S)--LogS can be well described by a power-law model and their maximum
likelihood best fit parameters have been reported in Table 5.

At the 4.5--7.5 keV flux limit of $f_x\geq \sim 7 \times 10^{-14}$ erg
cm$^{-2}$ the surface densities of optically type 1 AGN and type 2 AGN are $1.63\pm 0.25$
deg$^{-2}$ and $0.83\pm 0.18$ deg$^{-2}$, respectively.  Optically type 2 AGN represent
$34 \pm 9$\% of the AGN population shining in the  4.5--7.5 keV energy
selection band at the flux limit of the HBSS.
In Table 8 we compare the optically type 2 AGN fraction from the HBSS sample with  that
found in a few representative X-ray surveys. As expected from the CXB  synthesis
models the fraction of optically type 2 AGN decreases going to softer  X-ray surveys
(e.g. by a factor  $\sim 2.6$ in the XMM BSS survey and by a factor $\sim 34$
in the  ROSAT Bright Survey) while remaining almost constant (around 35\%) for
surveys  above 2 keV down to a flux limit of $10^{-14}$  \ecs.  In the flux
range between  $ 10^{-15}$ and  $10^{-14}$ \ecs there is apparently an increase
of the fraction of  optically type 2 AGN but the spectroscopic identification  rate in
this flux range is still low ($\sim 50\%$) preventing us  from speculating 
further.

\begin{table*}
\begin{center}
\caption{Optically Type 2 AGN fraction in few representative X-ray surveys.}
\begin{tabular}{lrrrr}
\hline
\hline
Sample                     & Energy Range   & Flux Range               & Type 2 Fraction      & Ref./Note    \\
(1)                        & (2)            & (3)                      & (4)                  & (5)          \\
\hline

ROSAT Bright Survey          & 0.5--2.0      & $> \sim 2\times 10^{-12}$ &  $1 \pm 0.5$\%       & (a)      \\
XMM BSS                      & 0.5--4.5       & $> 7\times 10^{-14}$     &  $13 \pm 4$\%        & (b)      \\
XMM HBSS                     & 4.5--7.5       & $> 7\times 10^{-14}$     &  $34 \pm 9$\%        & (c)      \\
HELLAS2XMM + Other Surveys   & 2--10          & $ 10^{-15} - 10^{-14}$   &  $57 \pm 12$\%       & (d)      \\
HELLAS2XMM + Other Surveys   & 2--10          & $ 10^{-14} - 10^{-13}$   &  $37 \pm 6$\%        & (e)      \\
\hline \hline
\end{tabular}
\end{center}
Columns are as follows:
(1) Sample;
(2) Energy Selection Band; 
(3) Investigated flux range;
(4) Type 2 AGN fraction over the total AGN density in the specific survey;
(5) References and Note as follows:
    (a) From Schwope et al., 2000;
    (b) From this paper using the BSS sources with Right Ascension below $5^h$ or 
        above $17^h$. For this BSS subsample the spectroscopic identification rate is 
	around $88\%$;
    (c) From this paper using the HBSS sample and taking into account the small 
        number of sources still unidentified as discussed in section 3.6;
    (d) Fraction of optically obscured AGN from Fiore
        et al., 2003. The spectroscopic ID rate of the sample in this flux range is 
	$\sim 50\%$;
    (e) Fraction of optically obscured AGN from Fiore
        et al., 2003. The spectroscopic ID rate of the sample in this flux range is 
	$\sim 81\%$.
\end{table*}

The optically type 2 AGN fraction from the HBSS sample is also in very good
agreement  with the fraction of X-ray absorbed ($N_H > 10^{22}$ cm$^{-2}$) AGN
found from  several studies (e.g. Piconcelli et al., 2003; 
Perola et al., 2004) which ranges between 25\% to 40\% for hard X-ray
fluxes spanning  four orders of magnitude, from  $10^{-10}$ to   $10^{-14}$
\ecs.  While we anticipate that the ``Modified Unification Scheme´´ for the 
synthesis of the CXB (Ueda et al., 2003) predicts a fraction  of absorbed AGN
around 41\% for the fluxes (and energy band)  covered by the HBSS, a proper
comparison will have to wait for 
the completion of  the X-ray spectroscopic work.   

\begin{figure}
\begin{center}
\includegraphics[width=0.50\textwidth]{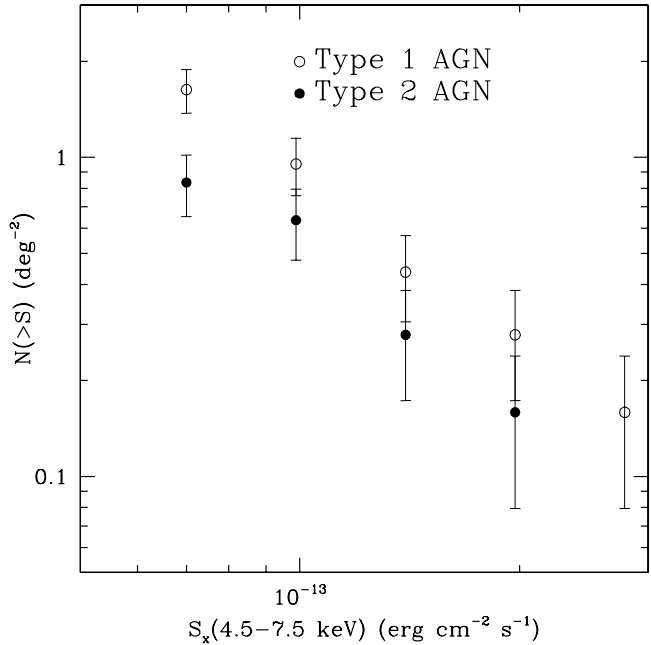}
\caption{ The number-flux relationship in the 4.5-7.5
keV energy band for optically type 1 (open circles) and type 2 
(filled circles) AGN obtained using the sources in the HBSS sample.
}
\end{center}
\end{figure}

\section {Summary and Conclusions}

In this paper we have discussed the scientific goals and the strategy of the
``The XMM-Newton Bright Serendipitous Survey". This survey comprises two
flux-limited samples: the BSS sample and the HBSS sample having a flux limit of
$\sim 7 \times 10^{-14}$ erg cm$^{-2}$ in the 0.5-4.5 keV and 4.5-7.5 keV
energy band, respectively.

From the analysis of 237 suitable XMM-Newton fields (211 for the
HBSS), covering a useful survey area of 28.10 (25.17 for the HBSS) sq. deg of
sky, we have defined (and presented here) a sample of 400 X-ray sources: 389
of them belong to the BSS, 67 to the HBSS with 56 X-ray sources in common. Up
to now $\sim 71$\% ($\sim 90$\%) of the sources in the BSS (HBSS) sample have
been spectroscopically identified, either from the literature or from our
spectroscopic observations.
The main results from this study can be so summarized:

a) the extragalactic number-flux relationship in the 0.5-4.5 keV and 4.5-7.5
keV band are in good agreement with previous and new results. They are well
described by a power law model, N($>$S) $\propto S^{-\alpha}$, with best fit
value for the slope $\alpha$ of $1.80 \pm 0.11$ and $1.64 \pm 0.24$ in the
0.5-4.5 keV and 4.5-7.5 keV bands, respectively;

b) at the X-ray flux limits of the survey presented here we found that the
optical counterpart in the  majority ($\sim 90$\%) of cases has an optical
magnitude brighter than the POSS II limit (R $\sim 21^{mag}$). Galactic counterparts 
represent about 14\% of the sources in the BSS sample and less than 3\%  of
the sources in the HBSS sample. The majority of the extragalactic objects
identified so far are broad line AGN both in the BSS ($\sim 80\%$)  and in the
HBSS ($\sim 60$\%);

c)  we have investigated the broad-band spectral properties of the selected
sources using  hardness ratios. No obvious trend of the source spectra as a
function of the count rate is measured. The average spectrum of the
``extragalactic" population corresponds to a (0.5-4.5 keV)
energy spectral index of
$\sim 0.8$ for the BSS sample and
$\sim 0.64$  for the HBSS sample. About
13\% (40\%) of the sources in the BSS (HBSS) sample seem to be described by an
energy spectral index flatter than that of the CXB; following the results
presented in Caccianiga et al., 2004 we speculate that these sources are
absorbed AGN with $N_H$ ranging from few times 10$^{21}$ up to few times 10$^{23}$
cm$^{-2}$.   There are hints from this study of a significant number of narrow
line AGN in the BSS sample occupying the locus typical of X-ray unabsorbed
broad line AGN.  A deeper investigation of their optical  and X-ray properties,
of the source selection function, 
as well as  a complete X-ray spectral analysis for all the BSS and HBSS sources
using data from all the EPIC  instruments (a major goal of this project) is in
progress  and will be presented elsewhere;

d) we do not find a compelling evidence that the HBSS 4.5-7.5 keV survey is
sampling a completely different source population compared with the BSS 0.5-4.5
keV survey.  Rather we find that the HBSS survey is  simply picking up a larger
fraction of absorbed AGN, consistent with  CXB synthesis models based on the
unification scheme of the AGN;

e) at the flux limit of the HBSS sample we measure surface densities of 
optically classified  type 1 and type 2 AGN of $1.63\pm 0.25$ deg$^{-2}$ and
$0.83\pm 0.18$ deg$^{-2}$, respectively. The AGN optically classified as  Type
2  represent $(34 \pm 9) \%$ of the total AGN population shining in the 4.5-7.5
keV energy band at the sampled  fluxes. 
A proper comparison with X-ray absorbed/unabsorbed AGN have to wait for 
the completion of the ongoing X-ray spectroscopic work.

f) finally, we found a clear separation between Galactic ``coronal emitting"
stars and extragalactic sources in the hardness ratio diagram and in the
(hardness ratio) vs. (X/O flux ratio) diagram. Since the investigated sample is a
fair representation of the high Galactic latitude X-ray sky, this result will
help with the selection of defined classes of sources from the XMM-Newton
catalogue prior to spectroscopic observations making the existing and incoming
XMM-Newton catalogs an unique resource for astrophysical studies.

\begin{acknowledgements}

This research has made use of the NASA/IPAC Extragalactic Database
(NED; which is operated by the Jet Propulsion Laboratory,
California Institute of Technology, under contract with the
National Aeronautics and Space Administration) and of the SIMBAD
database (operated at CDS, Strasbourg, France). We thanks Piero
Rosati for providing us some IDL routines used in the data
analysis and preparation.
We thanks L. Maraschi, G. Trinchieri, A. Wolter, D. Worrall,  N. Webb 
and the referee 
for a careful reading of the
paper and for useful comments which have improved the paper.
We thanks Y. Ueda to have provided us with the predictions from the 
Modified Unification Model in a tabular form.
RDC, TM, AC, PS, VB acknowledge partial
financial support by the Italian Space Agency (ASI grants:
I/R/037/01, I/R/062/02 and I/R/071/02), by the MURST
(Cofin-03-02-23) and INAF.
XB and FJC acknowledge financial support from
the Spanish Ministerio de Ciencia y Tecnolog\'{\i}a, 
under the project ESP2003-00812.
Based on observations collected at the Italian
``Telescopio Nazionale Galileo" (TNG), at the German-Spanish
Astronomical Center, Calar Alto (operated jointly by Max-Planck
Institut  f\"{u}r Astronomie and Instututo de Astrofisica de
Andalucia, CSIC), at the European Southern Observatory (ESO) and at the
Nordic Optical Telescope (NOT).
We would like to thank the staff members of the
TNG, ESO, Calar Alto and NOT for their support during the observations.
The TNG and NOT telescopes are operated in the island of La
Palma by the Nordic Optical Telescope Scientific Association and the Centro
Galileo Galilei of the INAF, respectively, in the Spanish Observatorio del Roque
de los Muchachos of the Instituto de Astrof\'{\i}sica de Canarias.
We finally thank the APM team for maintaining this facility.

\end{acknowledgements}

\Online

\setcounter {table} {1}

\begin{longtable}{lllrlllll}
\caption{Basic information on the XMM-Newton MOS2 fields used for the sample selection}\\
\hline
\hline
Obs. ID         & Filter  & RA; DEC (J2000)        & Exposure       & Log Nh    & $R_{in}$ & $R_{out}$ & BSS      & HBSS  \\
                &         &                        & s              & cm$^{-2}$ & arcmin   & arcmin    & srcs     & srcs   \\
(1)             & (2)     & (3)                    & (4)            & (5)       & (6)      & (7)       &  (8)     &  (9)      \\
\hline
\endfirsthead
\hline
\hline
Obs. ID         & Filter  & RA; DEC (J2000)        & Exposure       & Log Nh    & $R_{in}$ & $R_{out}$ & BSS     & HBSS  \\
                &         &                        & s              & cm$^{-2}$ & arcmin   & arcmin    & srcs    & srcs   \\
(1)             & (2)     & (3)                    & (4)            & (5)       & (6)      & (7)       & (8)     & (9)      \\
\hline
\endhead
\hline
\endfoot
0125310101 & Medium & 00 00 30.4 $-$25 06 43.4 &    19162.6 &         20.27 &          1 &         13 &  4 &  0 \\
0101040101 & Medium & 00 06 19.7 +20 12 22.8 &    34044.7 &         20.60 &          8 &         13 &  1 &  0 \\
0127110201 & Thin1  & 00 10 31.2 +10 58 40.7 &     7558.0 &         20.76 &          2 &         13 &  2 &  0 \\
0111000101 & Medium & 00 18 33.0 +16 26 08.0 &    31362.1 &         20.61 &          3 &         13 &  2 &  0 \\
0001930101 & Thin1  & 00 26 06.8 +10 41 12.5 &    18032.3 &         20.71 &          1 &         13 &  1 &  1 \\
0050140201 & Thin1  & 00 26 35.9 +17 09 37.8 &    50380.4 &         20.63 &          3 &         13 &  2 &  0 \\
0112320101 & Thin1  & 00 30 27.6 +04 51 47.4 &    24577.8 &         20.48 &          1 &         13 &  1 &  0 \\
0065770101 & Medium & 00 32 47.0 +39 34 33.3 &     7360.0 &         20.82 &          1 &         13 &  1 &  0 \\
\hline
\hline
\end{longtable}

Columns are as follows: (1) XMM-Newton Observation number; (2)
Filter used; (3) Right ascension and Declination (J2000) of the
MOS2 image center; (4) On-Axis good-time exposure; (5) Logarithm of the
Galactic hydrogen column density along the line of sight from
Dickey and Lockman, 1990; (6)
Inner radius of the part of the MOS2 image used; 
(7) Outer radius of the part of the MOS2 image used;
(8) total number of BSS sources found in the surveyed area of each MOS2 image;
(9) total number of HBSS sources found in the surveyed area of each MOS2 image. 
See section 2.4 for details.

NOTE -- $^{*}$ These 26 fields have not be considered for
 the definition of the HBSS sample.


\begin{landscape}
\begin{longtable}{lllrrrrl}
\caption{Basic information on the XMM-Newton BSS sample}\\
\hline
\hline
Name                 & Obs. ID    & RA; DEC (J2000)      & OffAxis  & Rate                     & HR2           & HR3            &  Class  \\
XBS...               &            &                      & arcmin     & $\times 10^{-2}$ cts/s    &               &              &      \\
(1)                  & (2)        & (3)                  & (4)       & (5)                       & (6)           & (7)            & (8)  \\
\hline
\endfirsthead
\hline
\hline
Name                 & Obs. ID    & RA; DEC (J2000)      & OffAxis  & Rate                     & HR2           & HR3            &  Class  \\
XBS...               &            &                      & arcmin     & $\times 10^{-2}$ cts/s    &               &              &      \\
(1)                  & (2)        & (3)                  & (4)       & (5)                       & (6)           & (7)            & (8)  \\
\hline
\endhead
\hline
\endfoot
J000027.7$-$250442 & 0125310101 & 00 00 27.8 $-$25 04 42.6 &  2.10 &$  1.13 \pm   0.08 $&$ -0.48 ^{+  0.06 }_{-  0.06 }$ &$ -0.37 ^{+  0.11 }_{-  0.11 }$& AGN1     \\
J000031.7$-$245502 & 0125310101 & 00 00 31.8 $-$24 55 02.9 & 11.68 &$  1.41 \pm   0.13 $&$ -0.67 ^{+  0.07 }_{-  0.06 }$ &$ -0.37 ^{+  0.17 }_{-  0.17 }$& AGN1     \\
J000100.2$-$250501$^d$ & 0125310101 & 00 01 00.2 $-$25 05 01.2 &  6.97 &$  1.38 \pm   0.10 $&$ -0.37 ^{+  0.07 }_{-  0.07 }$ &$ -0.47 ^{+  0.10 }_{-  0.10 }$& AGN1     \\
J000102.4$-$245850 & 0125310101 & 00 01 02.5 $-$24 58 50.7 & 10.72 &$  1.48 \pm   0.12 $&$ -0.54 ^{+  0.07 }_{-  0.07 }$ &$ -0.71 ^{+  0.11 }_{-  0.10 }$& AGN1     \\
J000532.7+200716 & 0101040101 & 00 05 32.8 +20 07 16.2 & 12.15 &$  2.47 \pm   0.12 $&$ -0.81 ^{+  0.03 }_{-  0.03 }$ &$ -0.70 ^{+  0.11 }_{-  0.10 }$& GAL        \\
J001002.4+110831 & 0127110201 & 00 10 02.4 +11 08 31.0 & 12.12 &$ 13.12 \pm   0.71 $&$ -0.79 ^{+  0.03 }_{-  0.03 }$ &$ -0.88 ^{+  0.07 }_{-  0.06 }$& star     \\
J001051.6+105140 & 0127110201 & 00 10 51.6 +10 51 40.2 &  8.61 &$  1.01 \pm   0.15 $&$ -0.76 ^{+  0.09 }_{-  0.10 }$ &$ -1.00 ^{+  0.76 }_{-  0.00 }$& star     \\
J001749.7+161952 & 0111000101 & 00 17 49.7 +16 19 52.5 & 12.11 &$  3.35 \pm   0.15 $&$ -1.00 ^{+  0.01 }_{-  0.00 }$ &$  1.00 ^{+  0.00 }_{-  0.96 }$& star     \\
\hline
\hline
\end{longtable}

Columns are as follows: (1)
Source name; (2) XMM-Newton Observation number; (3) Right Ascension and
Declination (J2000) of the source (X-ray position); (4) Angular
distance (in arcmin) between the source and the MOS2 image center;
(5) Source count rate, and $1\sigma$ error, in the 0.5-4.5 keV
energy band (units of $10^{-2}$ cts/s). In Table 1
we have reported the MOS2 conversion factors between the 0.5-4.5
keV count rate and the flux as a function of the energy spectral
index, the hardness ratio HR2 and the blocking filter; (6) and (7)
Hardness ratios computed as described in section 3.3. The errors
on the hardness ratios have been evaluated using simulations and
correspond to $1\sigma$; (8) Optical spectroscopic classification
(AGN1: broad line AGN; AGN2: narrow line AGN; GAL: Optically Normal
Galaxy; CL: Cluster of Galaxies; BL: BL Lac Object; star: star;
?: Tentative classification; see section 3.2 for details).

NOTE -- For two sources (XBS J052155.0$-$252220 and XBS
J213840.5$-$424241) the hardness ratio HR3 is undefined since these
sources are undetected above 2 keV.

$^a$ A detailed X-ray and optical spectral analysis of these sources have been
reported in Caccianiga et al. (2004).

$^b$ A detailed X-ray and optical spectral analysis of these sources have been
reported in Severgnini et al. (2003).

$^c$ These sources are more extended than the XMM-Newton EPIC MOS2 Point 
Spread Function at their off-axis angle.
Count rates have been evaluated using aperture photometry.

$^d$ A detailed X-ray spectral analysis of these sources have been
reported in Galbiati et al. (2004).

\end{landscape}

\begin{landscape}
\begin{longtable}{lllrrrrl}
\caption{Basic information on the XMM-Newton HBSS sample}\\
\hline
\hline
Name                 & Obs. ID    & RA; DEC (J2000)      & OffAxis  & Rate                     & HR2           & HR3            &  Class  \\
XBS...               &            &                      & arcmin     & $\times 10^{-3}$ cts/s    &               &              &      \\
(1)                  & (2)        & (3)                  & (4)       & (5)                       & (6)           & (7)            & (8)  \\
\hline
\endfirsthead
\hline
\hline
Name                 & Obs. ID    & RA; DEC (J2000)      & OffAxis  & Rate                     & HR2           & HR3            &  Class  \\
XBS...               &            &                      & arcmin     & $\times 10^{-3}$ cts/s    &               &              &      \\
(1)                  & (2)        & (3)                  & (4)       & (5)                       & (6)           & (7)            & (8)  \\
\hline
\endhead
\hline
\endfoot
J002618.5+105019$^b$ & 0001930101 & 00 26 18.5 +10 50 19.3 &           9.56 &$  2.35 \pm        0.55 $& -0.53 $^{+  0.04 }_{-  0.04 }$ & -0.67 $^{+  0.06 }_{-  0.06 }$& AGN1     \\
J013240.1$-$133307$^{b,e}$ & 0084230301 & 01 32 40.1 $-$13 33 07.8 &      11.90 &$  3.23 \pm    0.71 $& -0.02 $^{+  0.10 }_{-  0.10 }$ & -0.37 $^{+  0.11 }_{-  0.11 }$& AGN2     \\
J013944.0$-$674909$^b$ & 0032140401 & 01 39 44.0 $-$67 49 09.4 &       2.69 &$  2.05 \pm    0.46 $& -0.50 $^{+  0.07 }_{-  0.07 }$ & -0.43 $^{+  0.13 }_{-  0.13 }$& AGN1?    \\
J014100.6$-$675328$^b$ & 0032140401 & 01 41 00.7 $-$67 53 29.0 &       6.64 &$ 70.95 \pm    3.84 $& -0.45 $^{+  0.02 }_{-  0.02 }$ & -0.38 $^{+  0.03 }_{-  0.03 }$& star   \\
J015957.5+003309$^b$ & 0101640201 & 01 59 57.5 +00 33 09.7 &           9.56 &$  3.80 \pm        0.92 $& -0.63 $^{+  0.05 }_{-  0.05 }$ & -0.51 $^{+  0.10 }_{-  0.10 }$& AGN1     \\
J021640.7$-$044404$^{b,e}$ & 0112371701 & 02 16 40.7 $-$04 44 04.9 &       9.35 &$  2.08 \pm    0.49 $& -0.72 $^{+  0.04 }_{-  0.05 }$ & -0.26 $^{+  0.13 }_{-  0.13 }$& AGN1     \\
J021808.3$-$045845$^b$ & 0112371001 & 02 18 08.3 $-$04 58 45.7 &       2.26 &$  2.67 \pm    0.25 $& -0.65 $^{+  0.02 }_{-  0.02 }$ & -0.52 $^{+  0.04 }_{-  0.04 }$& AGN1     \\
J021817.4$-$045113$^b$ & 0112371001 & 02 18 17.4 $-$04 51 13.3 &       9.58 &$  3.30 \pm    0.40 $& -0.48 $^{+  0.02 }_{-  0.03 }$ & -0.56 $^{+  0.04 }_{-  0.04 }$& AGN1     \\
J021822.2$-$050615$^{a,b,d}$ & 0112371001 & 02 18 22.3 $-$05 06 15.7 & 8.48 &$ 4.54 \pm  0.43 $&  0.65 $^{+  0.09 }_{-  0.09 }$ &  0.40 $^{+  0.07 }_{-  0.07 }$& AGN2     \\
J023713.5$-$522734$^b$ & 0098810101 & 02 37 13.5 $-$52 27 34.4 &      12.07 &$  3.23 \pm    0.61 $& -0.57 $^{+  0.03 }_{-  0.03 }$ & -0.58 $^{+  0.07 }_{-  0.06 }$& AGN1    \\
J030206.8$-$000121$^b$ & 0041170101 & 03 02 06.9 $-$00 01 21.2 &      11.90 &$  3.10 \pm    0.42 $& -0.53 $^{+  0.03 }_{-  0.03 }$ & -0.46 $^{+  0.05 }_{-  0.06 }$& AGN1     \\
J030614.1$-$284019$^b$ & 0042340501 & 03 06 14.2 $-$28 40 19.9 &      10.87 &$  4.61 \pm    0.92 $& -0.49 $^{+  0.05 }_{-  0.05 }$ & -0.35 $^{+  0.09 }_{-  0.09 }$& AGN1     \\
J031015.5$-$765131$^b$ & 0122520201 & 03 10 15.6 $-$76 51 31.5 &       5.90 &$  4.39 \pm    0.46 $& -0.46 $^{+  0.03 }_{-  0.03 }$ & -0.53 $^{+  0.04 }_{-  0.04 }$& AGN1     \\
J031146.1$-$550702$^b$ & 0110970101 & 03 11 46.1 $-$55 07 02.5 &      12.38 &$  5.87 \pm    1.12 $& -0.58 $^{+  0.05 }_{-  0.05 }$ & -0.39 $^{+  0.11 }_{-  0.11 }$& AGN2    \\
J031859.2$-$441627$^{b,d}$ & 0105660101 & 03 18 59.3 $-$44 16 27.6 &  11.55 &$ 2.16 \pm   0.49 $& -0.21 $^{+  0.08 }_{-  0.08 }$ & -0.47 $^{+  0.10 }_{-  0.10 }$& AGN1     \\
J033845.7$-$352253$^{a,b}$ & 0055140101 & 03 38 45.8 $-$35 22 53.4 &  10.49 &$ 2.37 \pm   0.36 $& -0.05 $^{+  0.12 }_{-  0.12 }$ &  0.27 $^{+  0.11 }_{-  0.11 }$& AGN2     \\
J040658.8$-$712457$^{a,b}$ & 0111970301 & 04 06 58.9 $-$71 24 57.7 &  12.55 &$ 3.41 \pm   0.68 $&  0.39 $^{+  0.19 }_{-  0.19 }$ &  0.20 $^{+  0.16 }_{-  0.16 }$& AGN2    \\
J040758.9$-$712833$^{a,b}$ & 0111970301 & 04 07 59.0 $-$71 28 33.5 &  12.12 &$ 4.96 \pm   0.91 $&  0.49 $^{+  0.17 }_{-  0.17 }$ &  0.33 $^{+  0.13 }_{-  0.13 }$& AGN2     \\
J041108.1$-$711341$^b$ & 0111970301 & 04 11 08.1 $-$71 13 41.1 &      10.52 &$  2.20 \pm    0.42 $& -0.53 $^{+  0.08 }_{-  0.08 }$ & -0.04 $^{+  0.16 }_{-  0.16 }$& AGN1     \\
J050536.6$-$290050 & 0111160201 & 05 05 36.7 $-$29 00 50.8 &          12.43 &$  2.19 \pm        0.44 $& -0.27 $^{+  0.05 }_{-  0.05 }$ & -0.47 $^{+  0.07 }_{-  0.07 }$&          \\
J052108.5$-$251913$^e$ & 0085640101 & 05 21 08.5 $-$25 19 13.1 &           3.17 &$  2.11 \pm        0.55 $& -0.59 $^{+  0.05 }_{-  0.05 }$ & -0.55 $^{+  0.09 }_{-  0.09 }$& AGN1     \\
J052128.9$-$253032$^a$ & 0085640101 & 05 21 28.9 $-$25 30 32.4 &      10.73 &$  3.08 \pm    0.88 $&  1.00 $^{+  0.00 }_{-  0.17 }$ & -0.32 $^{+  0.15 }_{-  0.15 }$&          \\
J074202.7+742625 & 0123100101 & 07 42 02.7 +74 26 25.8 &              10.86 &$  3.38 \pm            0.49 $& -0.59 $^{+  0.03 }_{-  0.03 }$ & -0.28 $^{+  0.07 }_{-  0.07 }$& AGN1     \\
J074312.1+742937 & 0123100101 & 07 43 12.1 +74 29 37.4 &               5.33 &$ 10.92 \pm            0.64 $& -0.61 $^{+  0.01 }_{-  0.01 }$ & -0.56 $^{+  0.02 }_{-  0.02 }$& AGN1     \\
J080411.3+650906 & 0094400301 & 08 04 11.4 +65 09 06.2 &               9.49 &$  2.41 \pm            0.50 $&  0.06 $^{+  0.09 }_{-  0.09 }$ & -0.51 $^{+  0.09 }_{-  0.09 }$&          \\
J083737.1+254751 & 0025540301 & 08 37 37.2 +25 47 51.1 &              10.61 &$  7.30 \pm            1.12 $& -0.47 $^{+  0.04 }_{-  0.04 }$ & -0.53 $^{+  0.06 }_{-  0.06 }$& AGN1     \\
J083737.0+255151 & 0025540301 & 08 37 37.1 +25 51 51.2 &              12.25 &$  2.98 \pm            0.71 $& -0.34 $^{+  0.07 }_{-  0.08 }$ & -0.52 $^{+  0.10 }_{-  0.10 }$& AGN2     \\
J091828.4+513931 & 0084230601 & 09 18 28.5 +51 39 31.3 &               6.92 &$  3.03 \pm            0.54 $&  0.31 $^{+  0.09 }_{-  0.10 }$ & -0.39 $^{+  0.09 }_{-  0.09 }$& AGN1     \\
J095218.9$-$013643 & 0065790101 & 09 52 19.0 $-$01 36 43.1 &          11.78 &$ 24.50 \pm        2.98 $& -0.77 $^{+  0.03 }_{-  0.03 }$ &  0.41 $^{+  0.09 }_{-  0.09 }$& AGN1    \\
J101850.5+411506 & 0028740301 & 10 18 50.6 +41 15 06.6 &              10.32 &$  2.16 \pm            0.49 $& -0.69 $^{+  0.03 }_{-  0.03 }$ & -0.51 $^{+  0.07 }_{-  0.08 }$& AGN1     \\
J101922.6+412049 & 0028740301 & 10 19 22.6 +41 20 49.7 &               9.06 &$  2.46 \pm            0.47 $& -0.44 $^{+  0.04 }_{-  0.04 }$ & -0.58 $^{+  0.06 }_{-  0.05 }$& AGN1     \\
J104026.9+204542 & 0059800101 & 10 40 26.9 +20 45 43.0 &               9.32 &$  8.70 \pm            1.15 $& -0.58 $^{+  0.03 }_{-  0.03 }$ & -0.48 $^{+  0.05 }_{-  0.05 }$& AGN1     \\
J104522.1$-$012843 & 0125300101 & 10 45 22.1 $-$01 28 43.3 &          12.94 &$  3.00 \pm        0.67 $& -0.63 $^{+  0.04 }_{-  0.04 }$ & -0.46 $^{+  0.08 }_{-  0.08 }$& AGN1     \\
J104912.8+330459 & 0055990201 & 10 49 12.8 +33 04 59.8 &              10.46 &$  2.06 \pm            0.51 $& -0.45 $^{+  0.06 }_{-  0.06 }$ & -0.54 $^{+  0.09 }_{-  0.10 }$& AGN1?    \\
J110050.6$-$344331 & 0112880201 & 11 00 50.6 $-$34 43 32.0 &          12.71 &$  5.12 \pm        0.69 $& -0.48 $^{+  0.04 }_{-  0.04 }$ & -0.33 $^{+  0.07 }_{-  0.07 }$&          \\
J112026.7+431520$^a$ & 0107860201 & 11 20 26.7 +43 15 20.3 &           4.58 &$  2.32 \pm        0.37 $&  0.67 $^{+  0.08 }_{-  0.08 }$ & -0.32 $^{+  0.09 }_{-  0.09 }$& AGN2     \\
J113106.9+312518 & 0102040201 & 11 31 07.0 +31 25 18.2 &              11.18 &$  2.27 \pm            0.44 $& -0.53 $^{+  0.08 }_{-  0.08 }$ & -0.15 $^{+  0.14 }_{-  0.14 }$& AGN1     \\
J113121.8+310252 & 0102040201 & 11 31 21.8 +31 02 52.6 &              11.58 &$  3.88 \pm            0.69 $& -0.03 $^{+  0.07 }_{-  0.07 }$ & -0.35 $^{+  0.08 }_{-  0.08 }$& AGN2     \\
J113148.7+311358 & 0102040201 & 11 31 48.7 +31 13 58.8 &               8.45 &$  3.43 \pm            0.48 $&  0.10 $^{+  0.06 }_{-  0.07 }$ & -0.43 $^{+  0.07 }_{-  0.07 }$& AGN2      \\
J122656.5+013126 & 0110990201 & 12 26 56.6 +01 31 26.2 &               5.85 &$  3.07 \pm            0.54 $& -0.11 $^{+  0.09 }_{-  0.09 }$ & -0.34 $^{+  0.10 }_{-  0.10 }$& AGN2?    \\
J123600.7$-$395217 & 0006220201 & 12 36 00.7 $-$39 52 18.0 &           5.62 &$  2.66 \pm        0.29 $& -0.77 $^{+  0.01 }_{-  0.01 }$ & -0.77 $^{+  0.02 }_{-  0.02 }$& star     \\
J124641.8+022412 & 0051760101 & 12 46 41.8 +02 24 12.1 &               2.78 &$  2.22 \pm            0.49 $& -0.58 $^{+  0.05 }_{-  0.05 }$ & -0.48 $^{+  0.09 }_{-  0.09 }$& AGN1     \\
J132038.0+341124 & 0093640401 & 13 20 38.1 +34 11 24.3 &               3.10 &$  2.89 \pm            0.40 $& -0.43 $^{+  0.04 }_{-  0.04 }$ & -0.48 $^{+  0.06 }_{-  0.06 }$& AGN1     \\
J133942.6$-$315004 & 0035940301 & 13 39 42.6 $-$31 50 04.8 &          11.91 &$  3.52 \pm        0.51 $& -0.36 $^{+  0.05 }_{-  0.06 }$ & -0.20 $^{+  0.08 }_{-  0.08 }$& AGN1?    \\
J134656.7+580315$^a$ & 0112250201 & 13 46 56.7 +58 03 15.4 &          11.08 &$  3.33 \pm        0.56 $&  0.46 $^{+  0.10 }_{-  0.10 }$ & -0.09 $^{+  0.10 }_{-  0.10 }$& GAL        \\
J134749.9+582111 & 0112250201 & 13 47 49.9 +58 21 11.0 &               8.53 &$  7.39 \pm            0.66 $& -0.60 $^{+  0.02 }_{-  0.02 }$ & -0.54 $^{+  0.03 }_{-  0.03 }$& AGN1     \\
J140102.0$-$111224 & 0109910101 & 14 01 02.0 $-$11 12 24.3 &           9.27 &$  7.21 \pm        0.59 $& -0.52 $^{+  0.02 }_{-  0.02 }$ & -0.46 $^{+  0.03 }_{-  0.03 }$& AGN2?   \\
J140113.4+024016$^a$ & 0098010101 & 14 01 13.4 +02 40 17.0 &          12.78 &$  2.10 \pm        0.52 $& -0.55 $^{+  0.09 }_{-  0.09 }$ &  0.08 $^{+  0.16 }_{-  0.16 }$&          \\
J140127.7+025605 & 0098010101 & 14 01 27.7 +02 56 05.4 &               7.25 &$  6.66 \pm            0.62 $& -0.38 $^{+  0.02 }_{-  0.02 }$ & -0.53 $^{+  0.04 }_{-  0.04 }$& AGN1     \\
J141531.5+113156 & 0112250301 & 14 15 31.6 +11 31 56.2 &               4.15 &$  2.58 \pm            0.34 $& -0.55 $^{+  0.03 }_{-  0.03 }$ & -0.47 $^{+  0.06 }_{-  0.06 }$& AGN1     \\
J141830.5+251052$^c$ & 0109960101 & 14 18 30.5 +25 10 52.6 &               7.58 &$  7.70 \pm            1.20 $& -0.52 $^{+  0.04 }_{-  0.04 }$ & -0.65 $^{+  0.07 }_{-  0.07 }$& CL       \\
J142741.8+423335$^a$ & 0111850201 & 14 27 41.9 +42 33 35.8 &          11.45 &$  3.62 \pm        0.53 $&  0.48 $^{+  0.07 }_{-  0.07 }$ & -0.23 $^{+  0.07 }_{-  0.07 }$&          \\
J143835.1+642928 & 0111530101 & 14 38 35.1 +64 29 28.3 &              12.34 &$  3.57 \pm            0.54 $&  0.18 $^{+  0.07 }_{-  0.07 }$ & -0.37 $^{+  0.07 }_{-  0.06 }$& AGN1?    \\
J143911.2+640526$^a$ & 0111530101 & 14 39 11.2 +64 05 26.8 &          12.04 &$  2.44 \pm        0.39 $&  0.86 $^{+  0.09 }_{-  0.09 }$ &  0.29 $^{+  0.11 }_{-  0.11 }$& BL?     \\
J153452.3+013104$^e$ & 0112190401 & 15 34 52.4 +01 31 04.6 &              10.57 &$  8.36 \pm            1.30 $& -0.39 $^{+  0.04 }_{-  0.04 }$ & -0.55 $^{+  0.05 }_{-  0.05 }$& AGN1     \\
J160645.9+081525 & 0067340601 & 16 06 46.0 +08 15 25.1 &              12.61 &$  7.36 \pm            1.37 $&  0.81 $^{+  0.08 }_{-  0.08 }$ & -0.23 $^{+  0.10 }_{-  0.10 }$& AGN2?    \\
J161820.7+124116$^a$ & 0103461001 & 16 18 20.7 +12 41 16.3 &          11.56 &$  2.09 \pm        0.61 $&  0.11 $^{+  0.18 }_{-  0.18 }$ & -0.18 $^{+  0.18 }_{-  0.19 }$&          \\
J165425.3+142159 & 0113070101 & 16 54 25.4 +14 21 59.3 &               7.44 &$  5.26 \pm            1.08 $& -0.63 $^{+  0.03 }_{-  0.03 }$ & -0.61 $^{+  0.07 }_{-  0.07 }$& AGN1     \\
J165448.5+141311 & 0113070101 & 16 54 48.5 +14 13 11.6 &              12.71 &$  6.25 \pm            1.79 $& -0.54 $^{+  0.07 }_{-  0.07 }$ & -0.42 $^{+  0.13 }_{-  0.13 }$& AGN2     \\
J193248.8$-$723355$^b$ & 0081341001 & 19 32 48.8 $-$72 33 55.2 &       8.39 &$  4.47 \pm    0.73 $& -0.23 $^{+  0.08 }_{-  0.08 }$ & -0.08 $^{+  0.10 }_{-  0.10 }$& AGN2?    \\
J204043.4$-$004548$^b$ & 0111180201 & 20 40 43.5 $-$00 45 48.2 &      10.55 &$  3.24 \pm    0.80 $&  0.04 $^{+  0.11 }_{-  0.11 }$ & -0.35 $^{+  0.12 }_{-  0.12 }$& AGN2     \\
J205635.7$-$044717$^b$ & 0112190601 & 20 56 35.8 $-$04 47 17.9 &       9.96 &$  2.08 \pm    0.50 $& -0.65 $^{+  0.04 }_{-  0.05 }$ & -0.48 $^{+  0.11 }_{-  0.11 }$& AGN1     \\
J205829.9$-$423634$^b$ & 0081340401 & 20 58 30.0 $-$42 36 35.0 &       2.42 &$  3.91 \pm    0.56 $& -0.57 $^{+  0.03 }_{-  0.03 }$ & -0.47 $^{+  0.06 }_{-  0.06 }$& AGN1    \\
J213002.3$-$153414$^b$ & 0103060101 & 21 30 02.3 $-$15 34 14.1 &      12.83 &$  2.30 \pm    0.47 $& -0.65 $^{+  0.04 }_{-  0.04 }$ & -0.50 $^{+  0.09 }_{-  0.09 }$& AGN1     \\
J213820.2$-$142536     & 0092850201 & 21 38 20.2 $-$14 25 37.0 &      11.17 &$  2.80 \pm    0.59 $& -0.29 $^{+  0.07 }_{-  0.07 }$ & -0.45 $^{+  0.10 }_{-  0.10 }$& AGN1     \\
J214041.4$-$234720$^b$ & 0008830101 & 21 40 41.5 $-$23 47 20.1 &       9.80 &$  3.30 \pm    0.68 $& -0.57 $^{+  0.05 }_{-  0.05 }$ & -0.39 $^{+  0.10 }_{-  0.10 }$& AGN1     \\
J220601.5$-$015346$^b$ & 0012440301 & 22 06 01.5 $-$01 53 47.0 &      12.92 &$  2.26 \pm    0.55 $& -0.41 $^{+  0.06 }_{-  0.06 }$ & -0.41 $^{+  0.09 }_{-  0.09 }$& AGN1     \\
\hline
\hline
\end{longtable}
\end{landscape}

\newpage

Columns are as follows: (1) Source name;
(2) XMM-Newton Observation number; (3) Right Ascension and Declination
(J2000) of the source (X-ray position); (4) Angular distance (in
arcmin) between the source and the MOS2 image center; (5) Source
count rate, and $1\sigma$ error, in the 4.5-7.5 keV energy band
(units of $10^{-3}$ cts/s). In Table 1 we have
reported the MOS2 conversion factors between the 4.5-7.5 keV count
rate and the flux as a function of the energy spectral index, the
hardness ratio HR2 and the blocking filter; (6) and (7) Hardness
ratios computed as described in section 3.3. The errors on the
hardness ratios have been evaluated using simulations and
correspond to $1\sigma$; (8) Optical spectroscopic classification
(AGN1: broad line AGN; AGN2: narrow line AGN; GAL: Optically Normal
Galaxy; CL: Cluster of Galaxies; BL: BL Lac Object; star: star;
?: Tentative classification; see section 3.2 for details).

NOTE -- $^a$ These 11 sources belong to the HBSS sample but not to
the BSS sample, while the remaining 56 HBSS sources are present
also in the BSS sample.

$^b$ A detailed X-ray and optical spectral analysis of these
sources have been reported in Caccianiga et al. (2004).

$^c$ These sources are more extended than the XMM-Newton EPIC MOS2 Point 
Spread Function at their off-axis angle.
Count rates have been evaluated using aperture photometry.

$^d$  A detailed X-ray and optical spectral analysis of these
sources have been reported in Severgnini et al. (2003).

$^e$  A detailed X-ray spectral analysis of these
sources have been reported in Galbiati et al. (2004).


\newpage

\appendix

\section {Illumination Factor}

The MOS2 cameras consist of a mosaic of 7 identical,
front-illuminated CCDs with a dead space between the different
chips. It is clear that serendipitous X-ray sources falling close
to the gaps between the CCDs (or to their edges) could have either
the flux and/or the source centroid poorly determined. The poor
understanding of the corrections to be applied to these sources
could represent a problem  in the subsequent analysis and/or
interpretation of the data.
It order to take into account this problem in a objective way we have used
the procedure detailed below.

From the exposure map produced from the pipeline processing system
we have built a mask image representing the area on the sky which is
``effectively" imaged by the CCDs. To produce this
mask we have used
the SAS task {\it emask} with {\it threshold1}=0.25 and
{\it threshold2}=0.20 (see
http://xmm.vilspa.esa.es/external/xmm\_sw\_cal/sas\_frame.shtml for
specific details).

Using this mask image we have thus defined the ``Illumination
Factor"  of each source as the fraction of sky ``effectively"
imaged by the CCDs in a circle of 20 arcsec around the
source.

The ``Illumination Factor"  so defined ranges between 0.3 and 1 and
the lower the ``Illumination Factor" the closer is the source to
gaps and/or edges in the CCDs. In the BSS and/or HBSS catalogues we have
retained only sources with ``Illumination Factor"  $\gae 0.8$.
Given the PSF of the MOS2 detector and its
energy and off-axis dependence we have evaluated that less than
10\% of the flux is lost in the case of a source with an
``Illumination Factor"  equal to the lower limit of 0.8.
In table B.1 we report the complete
list of sources which meet the selection criteria for the BSS
and/or HBSS samples (e.g., inside the selected area
between the inner and outer radius of each MOS2 image, count rate
and likelihood limits, etc..) but that have been excluded from the
sample because their `Illumination Factor"  is below 0.8.

Finally the produced mask has been also used, in the
computation of the sky  coverage, to take into account
the excluded area because of edges and gaps.

\begin{table}
\begin{center}
\caption{Basic information on the sources excluded from the
sample(s) since their ``Illumination Factor" is less than 0.8}
\begin{tabular}{llrl}
\hline
\hline
Obs. ID    & RA; DEC (J2000)      & OffAxis  & Illum. \\
           &                      & arcmin    &    \\
(1)                  & (2)        & (3)       & (4) \\
\hline
 0111000101 & 00 18 41.6   +16 20 33.4  &   5.94     & 0.69 \\
 0001930101 & 00 26 24.9   +10 31 23.7  &  10.77     & 0.68 \\
 0112320101 & 00 30 08.3   +05 03 40.9  &  12.83     & 0.70 \\
 0065770101 & 00 32 41.0   +39 40 08.2  &   5.70     & 0.42 \\
 0112600601 & 01 26 40.3   +19 12 13.0  &  12.36     & 0.60 \\
 0112630201 & 01 34 03.1 $-$40 02 22.8  &  12.38     & 0.68 \\
 0112371501 & 02 18 21.9 $-$04 34 51.2  &   7.80     & 0.48 \\
 0111110501 & 02 22 49.4 $-$05 14 53.0  &   5.58     & 0.27 \\
 0098810101 & 02 36 30.3 $-$52 27 04.3  &   7.66     & 0.77 \\
 0098810101 & 02 37 02.1 $-$52 23 48.0  &   8.53     & 0.62 \\
 0122520201 & 03 13 14.6 $-$76 55 55.4  &   5.88     & 0.71 \\
 0110970401 & 03 13 34.3 $-$55 26 43.6  &   7.85     & 0.76 \\
 0111970301 & 04 09 02.1 $-$71 07 54.6  &   9.83     & 0.63 \\
 0085640101 & 05 21 00.2 $-$25 28 52.5  &   7.00     & 0.49 \\
 0050150101 & 05 25 57.1 $-$33 44 35.7  &  10.72     & 0.74 \\
 0110930101 & 06 17 56.8   +78 16 05.3  &   5.81     & 0.66 \\
 0103860101 & 06 24 46.8 $-$64 33 46.0  &  10.80     & 0.78 \\
 0112980201 & 06 57 26.2 $-$55 49 53.9  &  10.49     & 0.78 \\
 0025540301 & 08 38 27.2   +25 50 53.7  &   5.75     & 0.65 \\
 0025540301 & 08 38 52.5   +25 37 25.3  &  10.27     & 0.58 \\
 0110660201 & 09 07 51.1   +62 01 58.3  &  12.77     & 0.65 \\
 0084230601 & 09 16 45.2   +51 41 45.0  &  10.53     & 0.57 \\
 0110930201 & 10 00 56.7   +55 41 01.2  &   8.50     & 0.69 \\
 0101040301 & 10 23 13.3   +19 56 50.5  &   6.43     & 0.62 \\
 0055990201 & 10 50 29.0   +33 00 42.8  &   7.63     & 0.70 \\
 0110660401 & 11 26 11.9   +42 52 45.9  &   2.09     & 0.78$^{a}$ \\
 0124110101 & 12 21 34.3   +75 09 18.1  &   9.34     & 0.60 \\
 0124110101 & 12 22 07.3   +75 26 20.7  &   7.88     & 0.53 \\
 0092360601 & 12 52 14.1 $-$83 46 42.6  &   9.48     & 0.66 \\
 0002940101 & 13 06 32.8 $-$23 31 13.1  &  12.05     & 0.62 \\
 0111160101 & 13 34 16.9   +50 23 09.6  &   7.78     & 0.64 \\
 0111160101 & 13 35 09.6   +50 39 17.8  &  11.58     & 0.73 \\
 0111570201 & 13 35 18.7 $-$34 21 45.5  &   8.30     & 0.62 \\
 0112250301 & 14 15 39.9   +11 24 04.4  &   5.69     & 0.37 \\
 0109960101 & 14 18 42.7   +25 07 09.5  &   9.90     & 0.79 \\
 0070740301 & 15 03 39.6   +10 16 04.9  &  11.46     & 0.66 \\
 0018741001 & 15 18 59.7   +06 18 39.4  &   5.57     & 0.73 \\
 0061940301 & 16 32 39.0   +78 11 54.3  &   5.48     & 0.56 \\
 0102040101 & 17 22 54.3   +34 27 25.3  &  10.89     & 0.58 \\
 0061940201 & 21 37 47.7 $-$42 26 14.5  &  10.55     & 0.68 \\
 0008830101 & 21 39 57.9 $-$23 45 35.1  &   7.27     & 0.76 \\
 0111790101 & 22 36 06.1 $-$26 08 04.6  &   6.80     & 0.55 \\
 0109070401 & 22 47 48.1 $-$51 10 18.1  &   8.38     & 0.73 \\
 0025541001 & 23 04 43.5   +12 12 10.2  &   8.04     & 0.53 \\
 0123900101 & 23 14 32.2 $-$42 33 01.5  &  12.11     & 0.72 \\
\hline \hline
\end{tabular}
\end{center}
Columns are as follows:
(1) XMM-Newton Observation number; (2) Right Ascension and Declination
(J2000) of the source (X-ray position); (3) Angular distance (in
arcmin) between the source and the MOS2 image center; (4)
``Illumination factor".

NOTE -- $^{a}$Although this source is well contained inside the
MOS2 central CCD it fall above a bad column; for this reason its
``Illumination factor" is below 0.8.
\end{table}

\section {Cleaning procedure}

XMM-Newton observations are subject to ``flares" in the background
rate, probably due to soft protons  which are
collimated by the X-ray mirrors toward the EPIC cameras and
interact with the structure of the detectors and the detectors
itself. The current understanding is that soft protons are
probably organized in clouds populating the Earth's magnetosphere.

In order to check the background quality of the dataset used we
have defined a ``Background Estimator Parameter" which is roughly
proportional to the ``real background" accumulated in the MOS2
images.

To set this ``Background Estimator Parameter" for each image we
have produced an histogram of the total accumulated MOS2 counts in
the 10-12 keV energy range as a function of time; the histogram
bin size  has been set to 100 seconds.

Thus, using this histogram we have:

a) evaluated the mean count rate ($<bck>$) and its standard deviation
($\sigma_{bck}$)

b) eliminated the time intervals which have a count rate greater than
$<bck> + 2\times \sigma_{bck}$

c) repeated points a) and b) 10 times

The mean count rate at the end of the loop described above is the
``Background Estimator Parameter".

\end{document}